\documentstyle[aps,graphics,epsfig]{revtex}

\newcommand{\snnf}[0]{$\sigma_{nn}^{free}$~}
\newcommand{\snn}[0]{$\sigma_{nn}$~}
\newcommand{\Kinf}[0]{K$_{\infty}$~}

\newcommand{\Yr}[0]{Y$_{\rm r}$~}
\newcommand{\yr}[0]{Y_{\rm r}~}
\newcommand{\Ycm}[0]{Y$_{\rm cm}$~}

\newcommand{\yrcm}[0]{Y_{\rm rcm}~}
\newcommand{\Yproj}[0]{Y$_{\rm proj}$~}
\newcommand{\Ynn}[0]{Y$_{\rm nn}$~}
\newcommand{\ynn}[0]{$Y_{\rm rnn}$~}
\newcommand{\bexp}[0]{b$_{exp}$~}

\newcommand{\Ebal}[0]{E$_{\rm bal}$~}

\newcommand{\pxp}[0]{P$_{x}$~}

\newcommand{\pxpsa}[0]{$<$\pxp/A$>$~}

\newcommand{\Fexp}[0]{$F_{exp}$~}
\newcommand{\Finit}[0]{$F_{init}$~}

\newcommand{\degre}[0]{$^\circ$}

\newcommand{\figart}[5]{\begin{figure}[#1]
                        \begin{center}
                        \includegraphics*[scale=#5,draft=false]{#2.eps}
                        \caption{#3}
                        \label{#4}
                        \end{center}
                        \end{figure}}

\newlength{\larbox}

\newcommand{\figartdeuxcol}[7]
{
\larbox=0.49\textwidth
\begin{center}
\begin{figure}[#1]
\parbox[htbp]{\larbox}{
\includegraphics*[scale=#3,draft=false]{#2.eps}}
\parbox[htbp]{\larbox}{
\includegraphics*[scale=#5,draft=false]{#4.eps}}
\caption{\label{#7} {#6}}
\end{figure}
\end{center}
}

\newcommand{\figartdeuxcolvar}[9]
{
\larbox=0.49\textwidth
\begin{center}
\begin{figure}[#1]
\parbox[htbp]{#8\textwidth}{
\includegraphics*[height=#3,draft=false]{#2.eps}}
\parbox[htbp]{#9\textwidth}{
\includegraphics*[height=#5,draft=false]{#4.eps}}
\caption{\label{#7} {#6}}
\end{figure}
\end{center}
}
\newcommand{\LPC}[0]{LPC Caen
(IN2P3-CNRS/ISMRA et Universit\'e),
14050 Caen Cedex , France}
\newcommand{\SACLAY}[0]{DAPNIA-SPhN, CEA/Saclay,
91191 Gif-sur-Yvette Cedex, France}
\newcommand{\SUBATECH}[0]{SUBATECH (IN2P3-CNRS/Universit\'e),
44070 Nantes Cedex, France }
\newcommand{\GANIL}[0]{GANIL (DSM-CEA/IN2P3-CNRS),
B.P. 5027, 14076 Caen Cedex 5, France}
\newcommand{\IPNO}[0]{IPN Orsay (IN2P3-CNRS),
91406 Orsay Cedex, France}
\newcommand{\IPNL}[0]{IPN Lyon (IN2P3-CNRS/Universit\'e),
69622 Villeurbanne Cedex, France}
\newcommand{\IFIN}[0]{Nuclear Institute for Physics and Nuclear Engineering,
Bucharest, Romania   }
\newcommand{\NAP}[0]{Dipartimento di Scienze Fisiche, Univ. di Napoli,180126 Napoli,
Italy.  }
\newcommand{\CNAM}[0]{Centre National des Arts et M\'etiers, 
F-75141 Paris Cedex 03, France}

\newcommand{\modif}[1]{{ #1}}
\newcommand{\mod}[1]{{ #1}}

\graphicspath{{./figs_dphi/}{./fig_mod_2/}{./fig_mod/}{./figures/}}

\begin{document}
\title{\large \bf Measurements of sideward flow around the balance energy.}

\author{D.~Cussol~$^a$, T.~Lefort~$^a$, J.~P\'eter~$^a$, G.~Auger~$^b$, 
Ch.O.~Bacri~$^c$, F.~Bocage~$^a$, B.~Borderie~$^c$, R.~Bougault~$^a$, 
R.~Brou~$^a$, Ph.~Buchet~$^d$, J.L.~Charvet~$^d$, A.~Chbihi~$^b$, 
J.~Colin~$^a$, R.~Dayras~$^d$, A.~Demeyer~$^e$, D.~Dor\'e~$^d$, D.~Durand~$^a$, 
P.~Eudes~$^f$, 
E.~de~Filippo ~$^{d,}$\footnote{present address : INFN Corso Italia 57, 
95129 Catania, Italy}, J.D.~Frankland~$^c$, 
E.~Galichet~$^{e,i}$, E.~Genouin-Duhamel~$^a$, E.~Gerlic~$^e$, M.~Germain~$^f$, 
D.~Gourio ~$^{f,}$\footnote{present address : GSI, Postfach 110552, 64220 Darmstadt,
Germany}, D.~Guinet~$^e$, 
P.~Lautesse~$^e$, J.L.~Laville~$^b$, J.F.~Lecolley~$^a$, A.~Le F\`evre~$^b$, 
R.~Legrain ~$^{d,}$\footnote{Deceased}, N.~Le Neindre~$^a$, O.~Lopez~$^a$, 
M.~Louvel~$^a$, 
A.M.~Maskay~$^e$, L.~Nalpas~$^d$, A.D.~N'Guyen~$^a$, M.~Parlog~$^g$, 
E.~Plagnol~$^c$, G.~Politi, A.~Rahmani~$^f$, T.~Reposeur~$^f$, M.F.~Rivet~$^c$, 
E.~Rosato~$^h$, 
F.~Saint-Laurent ~$^{b,}$\footnote{present address : CEA, DRFC/STEP, 
CE Cadarache,
13108 Saint-Paul-lez-Durance, France}, 
S.~Salou~$^b$, 
J.C.~Steckmeyer~$^a$, M.~Stern~$^e$, G.~Tabacaru~$^g$, B.~Tamain~$^a$, 
L.~Tassan-Got~$^c$, O.~Tirel~$^b$, E.~Vient~$^a$, C.~Volant~$^d$, 
J.P.~Wieleczko~$^b$. (INDRA collaboration)\\~\\}

\address{ a) \LPC \\ b) \GANIL \\ c) \IPNO\\ d) \SACLAY \\ e) \IPNL\\
f) \SUBATECH \\ g) \IFIN \\ h) \NAP\\ i) \CNAM}


\maketitle
\begin{abstract}
Sideward flow values have been determined with the INDRA multidetector for 
{Ar+Ni, Ni+Ni and Xe+Sn} \mod{systems studied} at GANIL \mod{in the 30 to 100
A.MeV incident energy range}.
The balance energies found for { Ar+Ni and Ni+Ni} systems are in 
agreement with previous experimental results and theoretical
calculations. Negative sideward flow values have been measured. 
The possible origins of such negative values are discussed. They 
could result from \mod{a more important contribution of evaporated particles
with respect to the contribution {of} promptly 
emitted particles at mid-rapidity}. {But effects induced by the methods used 
to reconstruct the reaction plane cannot be totally excluded}. 
\mod{Complete tests} of these methods \mod{are} presented and the \mod{origins} 
of {the
``auto-correlation''} effect \mod{have been traced back}. For heavy fragments, 
the observed
negative flow values seem to be mainly due to the reaction plane reconstruction
methods. For light charged particles, these negative values could result from 
the \mod{dynamics} {of the collisions} and from the reaction plane 
reconstruction 
methods as well. These effects have to be {taken} 
into account when comparisons
with {theoretical calculations} are done.  
\end{abstract}



\section{Introduction}
Studies of sideward flow, also called in-plane flow, have been found to
provide {information} on the in-medium nucleon-nucleon
interaction. By comparing the experimental results {to dynamical 
calculations}, it is possible to {constrain} the value of the 
in-medium nucleon-nucleon cross section \snn, and the
incompressibility modulus of infinite nuclear matter \Kinf
\cite{Bertsch87,DLM92,West93,Shen93,Pop94,Ang97,Mag00}. The so-called 
balance energy \Ebal ({incident beam} energy for which the sideward flow
vanishes) has been found {to be} strongly 
{dependent} on \snn for light 
systems, and more {dependent} on \Kinf for heavier systems \cite{Mag00}.
\modif{For a fixed impact parameter and for a fixed incident energy, 
the flow parameter value strongly
depends on \Kinf \cite{DLM92,Molitoris87,Xu91}.} 
{A dependence on the total isospin of the system has been also
observed:} keeping the total mass constant, higher values of \Ebal are
{extracted} for the more neutron rich systems \cite{West98}. 

\modif{A simple interpretation of the sideward flow is that it 
results from first chance
nucleon-nucleon collisions. At incident energies below \Ebal, these collisions
are sentitive to the attractive part of the nucleon-nucleon interaction and
first chance particles are deflected toward an opposite direction relative to
the initial projectile direction. In this case, the flow parameter is negative.
At energies higher than \Ebal, the first chance collisions are sensitive to the
repulsive part of the nucleon-nucleon interaction and the first
chance particles are deflected toward the same direction relative to the initial
projectile direction. In that case the flow parameter is positive. At \Ebal, the
repulsive and the attractive part of the interaction counterbalance and 
first chance particles are not deflected: the flow parameter is zero.

But this scheme is indeed too simple.}
Dynamical calculations \cite{Ono95} have shown that the sideward flow 
{may result from} particles emitted at different {stages of
the reaction. Schematically,} one contribution {comes from} the 
decay of the so called quasi-projectile and quasi-target, {and the
other one, from the emission of particles in the first moments of the 
collision. The relative rate between these two contributions
is strongly dependent on the nature of the particle. This may explain} 
the different values of flow actually observed for different types of 
particles \cite{West93,Li96,Pak97}. 
\modif{In this more realistic frame, the study of the detailed 
evolution of the sideward flow with the
incident energy could shed a light on the production mechanism of particles
around the nucleon-nucleon velocity.}

The measurement of small sideward flow parameter values (typically
below 20 MeV/c/A) \mod{needs} high accuracy \mod{and a complete information on
each event which can be achieved} by using powerful 4$\pi$ 
multi-detectors. \modif{Experimentally only positive values of the flow 
parameter can be measured, since
the initial direction of the projectile is unknown, and since the positive
direction is defined as the mean direction of particles emitted above the
nucleon-nucleon frame velocity.} A full understanding of the experimental methods
is also needed, in order to correct possible spurious effects.

The aim of this paper is to
present the results of the sideward flow analyses {on Ar + Ni, 
Ni + Ni and Xe + Sn systems} \mod{from 25 to 95 A.MeV}, 
and to make an extensive test of the 
standard methods used
to measure the sideward flow. {Sideward flow measurements for 
the Ni + Ni 
system at high energies can be found in \cite{Reis98}.} In the
first section, the experimental set-up will be briefly described. 
The experimental
results will be presented in the second section. The third section will be
devoted to the test {of \mod{various} methods} used to reconstruct the reaction plane. 
Conclusions will be drawn in the last section. 

\section{Experimental setup}
The experiments were performed at the GANIL facility with the INDRA 
detector. {Target thicknesses were respectively 193 
$\rm{{\mu g/cm^2 ~of~^{58}Ni}}$ for the \modif{$^{40}$Ar + $^{58}$Ni} 
experiment, 179 
$\rm{{\mu g/cm^2 ~of~^{58}Ni}}$ for the \modif{$^{58}$Ni + $^{58}$Ni} 
experiment and 330 
$\rm{{\mu g/cm^2 ~of~^{nat}Sn}}$ for the \modif{$^{129}$Xe + $^{nat}$Sn} 
experiment.} 
Typical beam intensities were 3-4 ${\times 10^7}$ pps. 
A minimal bias trigger was used: events {were} 
registered when at least 
{three} charged particle detectors fired. 

The INDRA detector can be schematically described as a set of 17 detection 
rings centered on the beam axis. In each ring the detection of charged 
products was provided with two or three detection layers. The most forward 
ring, ${2^{\circ} \le \theta_{lab} \le 3^{\circ}}$, is made of phoswich 
detectors (plastic scintillators NE102 + NE115). Between ${3^{\circ}}$ 
and ${45^{\circ}}$ 
eight rings are constituted by three detector layers: ionization chambers, 
silicon and ICs(Tl). Beyond ${45^{\circ}}$, the eight remaining 
rings are made of double layers: ionization chambers and ICs(Tl). For the Ar + 
Ni experiment the ionization chambers {beyond $90^{\circ}$} 
were not yet installed. The total number of 
detection cells is 336 and the overall geometrical 
efficiency of INDRA detector corresponds to 90\% of 4${\pi}$. A complete 
technical description of the INDRA detector and of its electronics \mod{is} 
given in \cite{INDRA1,INDRA2} . Isotopic separation was achieved up to 
{Z=3-4} in the 
last layer (ICs(Tl)) over the whole angular range (${3^{\circ} \le 
\theta_{lab} \le 176^{\circ}}$). Charge identification was carried out up 
to \mod{Z=55} in the forward region (${3^{\circ}} \le \theta_{lab} \le 
\rm{45^{\circ}}$) and up to {Z=20} in the backward region 
(${\theta_{lab} \ge 45^{\circ}}$). The energy resolution is \mod{about}
5\% for ICs(Tl) and ionization chambers and better than 2\% for
Silicon {detectors}.

The INDRA detector \mod{capabilities} allow one to carry out an event by event 
analysis and to determine reliable global variables related 
to the impact parameter.

\section{Experimental results}
\subsection{\label{b}\textbf{Event sorting}}
The first step was to sort events as a function of the 
violence of the collision. In this paper, we will use the total 
transverse energy:

\begin{equation} 
 {E_{trans,tot}=\sum_{i=1}^M E_{k,lab}^i \times (sin({\theta_i}))^2}
\end{equation} 

where ${M}$ is the \mod{charged particle} multiplicity of the event, 
${E_{k,lab}^i}$, 
${\theta_i}$ {are respectively the kinetic energy 
and the polar angle (with respect to the beam)} 
\mod{of the particle $i$ in the laboratory frame}.
\modif{QMD calculations have shown that ${E_{trans,tot}}$ is a good
indicator of the true impact parameter at intermediate energies
\cite{SteckBormio95}.}

In order to sort events, we \mod{assumed}
a geometrical correspondence between ${E_{trans,tot}}$ 
and the impact parameter. { The} cross section of each bin \mod{was} expressed 
as an experimentally estimated impact parameter, ${b_{exp}}$ 
\cite{Pet90a}. 
The first bin in ${E_{trans,tot}}$, i.e the 31,4 mb of the highest 
values of ${E_{trans,tot}}$, is linked to ${b_{exp}}$ between 0 and 
1 fm, the second, to ${b_{exp}}$ between 1 and 2 fm, etc ...  The last 
bin corresponds to ${b_{exp}}$ larger than 8 fm. \modif{This procedure has been
applied on all detected events. Due to trigger conditions which remove the most
peripheral reactions, the estimate of the impact parameter is not very accurate
for the lowest values of ${E_{trans,tot}}$}. From now, $b_{exp}$ will refer
to the experimentally estimated impact parameter using the total transverse
energy. \modif{Previous studies
\cite{Ang97} have shown that the flow parameter values are weakly 
sensitive on the exact choice of the sorting variable.}

\subsection{\label{selec}\textbf{Selection of ``well measured'' events}}
The next step in the analysis was to select events in which sufficient 
information was \mod{recorded}. This was achieved by requiring that the total 
measured \mod{$\sum_{i=1}^{M}{Z_i \times V^i_{par}}$ 
(product of the charge $Z_i$ of particle $i$ 
{by}
its parallel velocity $V^i_{par}$) be larger than 70\% of the initial $Z_{proj} \times V_{proj}$ of 
the projectile \cite{Pet90a}}.

Since we want to study the dependence on impact parameter of in-plane flow, 
we checked that this selection conserves the \mod{whole impact 
parameter range}. {For all systems, we} 
noted that the total transverse energy 
distribution of selected events covers the whole range 
of total transverse energy of registered events. If one assumes that
the total transverse energy is a good {measurement} of the
violence of the collision, then this result indicates that the whole 
impact parameter range of registered events 
is kept in selected events. Indeed, the major part of eliminated events 
corresponds to some peripheral collisions in which the \mod{target like
fragment (TLF)} was not detected, the \mod{projectile like fragment (PLF)} 
was lost {in} the forward 0-2$^{\circ}$ beam hole and only
few light particles were detected. \mod{In this case, $E_{trans,tot}$ is still
correctly measured and since the TLF and PLF transverse
energies are very small.}

\subsection{\textbf{Flow parameters}}

\subsubsection{Definition}

To evaluate the flow parameters, one needs first to 
\mod{determine the reaction plane
on a event by event basis}. \mod{This is done by determining a
transverse axis which defines with the beam axis the reaction plane.}
The transverse 
momentum method \cite{Dan88} and the momentum tensor 
method \cite{Cugnon83} { have been} used. These methods have
been found to be equivalent. {As already mentioned in \cite{Ang97}, 
they give a better accuracy} on the 
reaction plane determination than the azimuthal correlations method 
{described in \cite{Wilson90}}. 
\mod{With such methods, the transverse axis is by definition oriented on the
mean transverse direction of the forward emitted products 
($V^i_{par} > V_{cm}$).} 
{Within the standard
interpretation of flow,} the measured parameters values should \mod{be 
positive}. {More details about these methods will be given
in section \ref{desc_meth}.}

In order to
avoid the so called ``auto-correlation'' effect, the particle of
interest {is usually} removed from the reaction plane determination, 
and a corrective momentum
{is} added to the momentum of other particles \cite{Dan88,Ogilvie89,Sul90}. 
{In this case, one plane per particle is determined.} 
The tests of the methods and the
issue of auto-correlations will be presented {in section \ref{TestMeth}}.

Once the reaction plane is determined, the projection of transverse momenta on
the reaction plane can be evaluated. For each reduced rapidity bin \Yr = Y /
\Yproj, its mean 
value \pxpsa is calculated, \modif{where
$Y=\frac{1}{2}log\left(\frac{1.+v_z/c}{1.-v_z/c}\right)$ 
is the rapidity, $v_z$ the parallel to the beam component of the velocity in the
laboratory frame and $c$ the light velocity}. 
The flow parameter $F$ \modif{is by definition the 
increase of \pxpsa from Y=\Ynn (\Yr=0.5) to
Y=\Yproj (\Yr=1) by using the slope of the function 
\pxpsa=f(\Yr) at mid-rapidity \cite{Sul90,Ogilvie90}. It reads}:

\begin{equation}
F =\frac{1}{2} \left(\frac{\partial \left< P_{x}/A \right>}
{\partial Y_r}\right)_{(Y_r=0.5)} 
\end{equation} 

{Typical evolutions of \pxpsa with \Yr are shown in figures
\ref{f:pxa_vs_yr_1} and \ref{f:pxa_vs_yr_2} for Ar + Ni collisions 
at 74 A.MeV. The rows correspond to the particle types, the columns, to the 
experimental impact parameter bins. The momentum tensor method has 
been used to reconstruct the reaction plane. For each panel, open 
circles correspond to one plane per particle (the particle of interest 
has been removed from the reaction plane reconstruction), full
circles, to one plane per event (all products are taken into account
for the reaction plane determination). For the Ar + Ni system, the balance
energy is expected around 80 A.MeV for the central collisions. Therefore, 
small flow parameter values should be measured at 74 A.MeV. 

For the one plane per event prescription (full circles), 
the slopes at mid-rapidity are very high and much
larger than the expected values. These overestimations are due to the 
so-called ``auto-correlation effect''. The flow values determined with the one 
plane per particle prescription (open circles) are lower and, as a result, in a
better agreement with the expected values. Some of them are 
even negative. Possible explanations of these negative slopes will be 
given in the following sections.} 

\subsubsection{Evolution of the flow parameters with the incident energy.}

For the most central collisions (\bexp $\le$ 3 fm), the results are
shown in figure \ref{f:flot_arni} for the Ar + Ni system, 
{in figure \ref{f:flot_nini} for the Ni + Ni system and in
figure \ref{f:flot_xesn} for the Xe + Sn system}. 
In {the first two} experiments, \mod{the reaction plane was determined by using
the momentum tensor method with the one plane per particle prescription.}
\modif{The flow parameter values has been determined by a linear fit in the \Yr
range from 0.4 to 0.6. The errors on flow parameter values 
have been estimated by changing the \Yr range from 0.35 to 0.65 and from 0.45 to
0.55. The error bar do not appear on the figures 
when it is smaller that the symbol size.}
 
The evolution of the flow
parameter with the incident
energy has a typical U shape for all { particles}. The incident energy which
corresponds to the minimum flow value is located {around 82 $\rm{\pm}$ 2 
A.MeV for the Ar + Ni system and around 75 $\rm{\pm}$ 2 A.MeV for the 
Ni + Ni system.} For both systems, these {balance energies do
not depend on the particle nature \mod{as for systems in reference \cite{West98}}. They
are in agreement with theoretical work \cite{DLM92}}. \mod{These calculation
were performed with \Kinf$\approx$~220 MeV and \snn=0.8~\snnf, where \snnf is the
free nucleon-nucleon cross-section}. 

{For the Xe + Sn system (figure \ref{f:flot_xesn}), the results are shown 
for two reaction plane reconstruction procedures: the momentum tensor 
method (open diamonds) and the transverse momentum method (stars) \mod{ with the
one plane per particle prescription}. 
As expected, both methods give close results. 
The U shape is barely seen for $^3$He and heavy particles 
(Z $\rm{\ge}$ 3). For all particle types, the minimum flow energy is 
difficult to determine, since the flow parameter
weakly depends on the incident energy. 
\modif{Unfortunately, the expected \Ebal value is around 50 A.MeV which is the
maximum incident energy available for this system. No accurate determination of
\Ebal can be done for this system}. The \mod{full} circles correspond to the momentum tensor 
method when one plane per event is determined. The flow 
parameter values are higher than the value obtained with one plane per 
particle due to the ``auto-correlation'' effect. 
The same behavior is observed with the lighter systems 
Ar + Ni and Ni + Ni when one plane per event is reconstructed.} 

But {the most striking feature of figures \ref{f:flot_arni},
\ref{f:flot_nini} { and \ref{f:flot_xesn} \mod{for the one plane per particle
prescription}} is the observation of 
negative flow parameter values for d, t,
$^3$He and fragments with a charge greater than 3. 
\mod{By definition only positive values are expected}. 
\mod{Negative} flow parameter values have 
already been
observed in previous studies \cite{Krof92}\mod{, but} 
no clear explanation was given for
this effect. We will propose in the {next two} sections two 
possible scenarios for these negative flow values.

\section{The physical effect for negative flow values.}

A possible explanation for the negative flow values is given by {AMD 
calculations \cite{Ono95}: for light charged particles, 
the flow of promptly emitted (named ``direct'') particles can
be opposite to the flow of ``evaporated'' particles (emitted from the 
\mod{quasi-projectile} QP and the \mod{quasi-target} QT). By definition, 
the reaction plane is oriented positively 
in \mod{the mean direction of forward emitted products}. 
Therefore, negative flow values can be measured.}

The direction of the ``direct'' flow {in the reaction plane results
mainly from screening effects. As shown in figure \ref{f:flot_contribs}, 
in the first moments of the reaction, 
``direct'' nucleons coming from the projectile (target) are
screened by the target (projectile) nucleus. For a fixed impact
parameter, the orientation of the ``direct'' flow is determined by 
the geometrical
configuration at the touching point and is weakly dependent on the 
incident energy. It is always aligned on the same direction as the initial
orientation of the projectile in the reaction plane. At variance, the 
direction of the ``evaporated'' flow is aligned on the final
directions of the QP. This direction depends strongly on the incident 
energy, as it will been explained in the next paragraph.} Experimental 
studies have shown that heavy fragments could be emitted by non-evaporative 
processes, like a neck break-up
\cite{Bougault89,Stuttge92,Lecolley95,Larochelle99,Dempsey96,Toke95,Chen96,Montoya94,Lukasik97,Casini93,Stefanini95,Bocage}. 
Since the direction of emission of these fragments
is mainly aligned on the QP-QT axis, their flow direction is identical to
the so-called ``evaporated'' flow. The contribution of heavy fragments to
the flow parameter will be always attributed to the ``evaporated'' component.
If one assumes that the observed flow parameter results only from
these two contributions, {the evaporated and prompt emissions,} a
fairly simple picture can be proposed as shown in figures 
\ref{f:flot_contribs}. \modif{The simple pictures 
shown here are for pedagogical purpose. They do not pretend to reproduce
the true process. They are shown to give a feeling of what
kind of physical effect could lead to the observation of negative flow parameter
values.}

At low incident energy, where
the attractive part of the nucleon-nucleon interaction dominates, the QP
and the QT are deflected to the opposite side {relative to} their initial
directions (left column of figure \ref{f:flot_contribs}). In this
case, since \pxpsa is defined positive for particles which are
deflected on the same side as the QP, the variation of \pxpsa with 
\Yr {is} different for the two emissions: a ``S'' shape
for the ``evaporated'' particles {with a positive slope around
mid-rapidity} (dashed line in the lower left picture of figure
\ref{f:flot_contribs}) and a straight line with a negative slope for the
``direct'' particles (dashed and dotted line). 

At higher energies, where the
repulsive part on the interaction is dominant, the QP ``bounces'' 
\mod{on the QT} (right
column of figure \ref{f:flot_contribs}). In this case, the variations
of \pxpsa with \Yr {give a positive slope around mid-rapidity for the
two emissions as shown in the lower right panel of figure 
\ref{f:flot_contribs}.} The observed variation of \pxpsa with \Yr 
is a \mod{combination} of these two contributions with their associated \Yr
distributions. 

{Such \mod{combinations} are schematically shown in figure 
\ref{f:flot_neg}. For each side
panel, the original \pxpsa evolution of the ``direct'' particles
(straight dashed and dotted line) and its corresponding \Yr distribution 
(dashed and dotted gaussian distribution) are plotted. The \pxpsa contribution
(``S'' shape)
and the \Yr distribution of ``evaporated'' particles (sum of two gaussian
distributions) are plotted as dashed
lines. The resulting evolution of \pxpsa corresponds to the thick line. It is
simply obtained by summing the two \pxpsa contributions weighted by their
corresponding \Yr distributions.}

If the contribution of ``evaporated'' particles is dominant, then the
resulting flow parameter value is always positive (upper row of figure 
\ref{f:flot_neg}). {The ``S'' shape is weakly affected by the
contribution of ``direct'' emissions.} At variance, if ``direct''
emissions become dominant (lower row of figure \ref{f:flot_neg}), then
a negative flow value can be obtained. {At low incident energies, 
the evolution of \pxpsa with \Yr is dominated by the ``evaporated'' 
particles. When 
``direct'' emissions becomes dominant, the evolution of \pxpsa with \Yr follows
the one of the ``direct'' particles, and the ``S'' shape is strongly deformed.
Negative slopes can be found at mid-rapidity, i.e. negative flow parameter
values.} 

This explanation is very tempting for {the light charged particles. 
Deuterons, 
tritons and $^3He$ are predominantly emitted in the mid-rapidity
regions, whereas protons and alpha particles are emitted as much by
the QP and QT as by the mid-rapidity emissions
\cite{Lefort00,Dore98}. At the same time, positive flow values are
measured for alpha and protons and negative values for deuterons,
tritons and $^3He$ (see figures \ref{f:flot_arni}, \ref{f:flot_nini}
and \ref{f:flot_xesn}). Both experimental features support the \mod{above}
interpretation.} But this does not 
explain the negative values observed for the heavier fragments. \modif{Within
the simple interpretation of sideward flow, the emission of fragments via 
cluster-cluster} 
collisions of preexisting 
fragments \mod{in both partner bulk has a very low probability}. 
\mod{As already mentioned,} for PLF and the TLF, 
this observation is even in contradiction {to} 
the reaction plane
orientation which {is always} oriented along the direction of the
quasi-projectile. The solution of this puzzle has to be found elsewhere.

Previous studies \cite{Sul90} have shown that the reaction plane
determination method and the associated issue of auto-correlations could
strongly disturb the measurement of flow parameter. Before attributing the
observation of negative flow parameter values to physical effects,
one has { first} to be sure that these negative values are not due to the 
{analysis}
methods. The next section is devoted to the test of reaction plane
determination methods. \modif{The origins of the so called
{``auto-correlation''} effects will also be studied}.

\section{Test of the reaction plane determination methods}
\label{TestMeth}

\subsection{\textbf{Reaction plane estimation.}}
\label{desc_meth}

Let us now briefly describe three commonly {used methods}: the 
transverse momentum method \cite{Dan88}, the momentum tensor method
\cite{Cugnon83} and the azimuthal correlations method \cite{Wilson90}.
\modif{From now on, the $Oz$ axis is the beam axis, the $Ox$ axis is the axis
in the reaction plane  which is perpendicular to the beam axis and $Oy$ 
axis is the axis perpendicular to the reaction plane.}
The transverse momentum method, explained in details in \cite{Dan88}, 
is {based} on the fact that the sum of
transverse momenta of particles emitted by the quasi-projectile
$\overrightarrow{P_{QP}}$ is opposite to
the sum of the transverse momenta of particles emitted from the quasi-target 
$\overrightarrow{P_{QT}}$. \mod{This is valid within the binary mechanism
hypothesis, where the mid-rapidity contribution is negligible}. 
Those vectors belong to the reaction plane. In
order to maximize the efficiency of the method, one has to calculate \mod{the}
difference of the two vectors $\overrightarrow{Q}=\overrightarrow{P_{QP}}-
\overrightarrow{P_{QT}}$. Usually, $\overrightarrow{Q}$ is determined \mod{in} 
the following way:
\begin{equation}
\overrightarrow{Q}=\sum_{j=1}^{N}{\omega_j \overrightarrow{P_j^\perp}}
\end{equation} 
\noindent where $N$ is the total number of {particles} in the event,
$\overrightarrow{P_j^\perp}$ is the transverse momentum of
particle $j$ and $\omega_j$ a weight defined as follows:
\begin{equation}
\omega(\yr)=\left\{
\begin{array}{rrr}
 -1& \mbox{if} &\yr-\yrcm < -\delta \\
\\
 0& \mbox{if} &-\delta \leq \yr-\yrcm \leq \delta \\
\\
 1& \mbox{if}& \yr-\yrcm > \delta
\end{array}
\right.
\end{equation}
\vskip 0.5 cm
\noindent or
\begin{equation}
\omega(\yr)=\yr-\yrcm
\end{equation}
\noindent where $\yr$ is the reduced rapidity of the particle, $\yrcm$ the
center of mass reduced rapidity (close to the reduced rapidity of the
nucleon-nucleon frame \ynn for the symmetric systems) 
and $\delta$ a parameter which {allows} 
to remove
{ the mid-rapidity particles from the estimation of the reaction plane}. 
$\overrightarrow{Q}$ defines
with the beam direction the reaction plane. With this definition, 
the reaction plane is systematically
oriented along the quasi-projectile direction.  

Another way to estimate the reaction plane is to calculate and diagonalize 
\mod{a} tensor $T_{\mu,\nu}$. \mod{The beam axis defines with the
eigenvector corresponding to the {highest} eigenvalue}  
the reaction plane. As for the transverse momentum the reaction plane is
oriented along the direction of the quasi-projectile. The tensor is defined {
in} the following way:

\begin{equation}
T_{\mu\nu}=\sum_{j=1}^{N}{\Omega_j \times P_j^\mu \times P_j^\nu} 
\quad\mbox{with}\quad \mu,\nu = x,y,z
\end{equation}

\noindent where $P_j^\mu$ is the momentum component along the $\mu$ 
axis ($\mu = x,y,z$) for { the} particle $j$. $\Omega_j$ is a weight
which is usually set to {the inverse of the mass} $A_j$ 
of the particle {$\Omega_j = 1/A_j$} \mod{(energy tensor)}.

The azimuthal correlations method is based on the following observation: 
in case of strong
in-plane emission (high flow parameter value), the sum of the distances 
of particles momenta with respect to that plane are minimum \cite{Wilson90}. 
This sum $D^2$ is calculated as follows:
\begin{equation}
D^2=\sum_{j=1}^{N} {\left[ \left(P_j^x\right)^2 + \left(P_j^y\right)^2 - 
\frac{\left(P_j^x+ a P_j^y\right)^2}{1+a^2}\right]} 
\end{equation}
\noindent where $P_j^x$ and $P_j^y$ are {the transverse} momentum
components along
the $Ox$ and $Oy$ axis respectively and $a=\tan (\varphi)$, where $\varphi$ is
the angle of the reaction plane relative to the $Ox$ axis. Experimentally, one
has to find the value of $\varphi$ which minimizes $D^2$. 
With this method, the orientation of the reaction plane is not defined.
One has to use the transverse momentum method to find it. In case of strong
out-of-plane particle emission ({ squeeze-out}), the estimated reaction plane
angle is wrong by $\pi / 2$. 

\subsection{\textbf{Testing procedure}}

Since the flow parameter is obtained from the average value \mod{of 
the transverse momentum projection} on the reaction plane, 
one has to rebuild this plane
from the experimental data. We have checked the \mod{reliabilities
of} the three \mod{reaction plane reconstruction} methods.
The general procedure of 
the test is the
following: a known flow parameter value $F$ is set for a sample of 
generated events;
then the reaction plane estimation method is applied on that sample and the so
called ``experimental flow parameter'' \Fexp is determined. A method is
considered effective if the experimental value \Fexp is equal or close to the 
initial one \Finit. This {allows} to \mod{also check} the \mod{additional} 
disturbance introduced by the
experimental set-up compared to the method itself.

To set the flow parameter value $F$ to an event, an in-plane component 
$P_x^a$ is added
to the transverse momentum of each particle, similarly to the procedure used in
\cite{Sul90}. The amplitude of this in-plane
component depends on the reduced rapidity of the particle: 
\begin{equation}
P_x^a(F,\yr)=
\left\{
\begin{array}{lcc}
-\frac{A \times F}{2}& \mbox{~if~} &\yr < 0.25 \\
\\
2 A F (\yr-0.5)& \mbox{~if~} &0.25 \leq \yr \leq 0.75 \\
\\
\frac{A \times F}{2}& \mbox{~if~}& \yr > 0.75
\end{array}
\right.
\end{equation}
\vskip 0.5 cm
\noindent where $F$ is the flow parameter value to be set on, \Yr is the
reduced rapidity of the particle, $A$ its mass and
$P_x^a(F,\yr)$ the in-plane component.

The test has been performed on two systems Ar + Ni at 74 A.MeV and Xe + Sn at
50 A.MeV. These two systems have been studied with INDRA and their comparison
{allows} to check {the effect of the mass of the system} on
the transverse flow measurement. \mod{These energies} values have been 
chosen because they are close to the expected
balance energy. For each system, 15000 events have been generated using the 
SIMON code, {whose} entrance channel includes a pre-equilibrium 
emission of protons and neutrons \cite{SIMON}. 
\modif{SIMON is not used to reproduce the experimental data, but rather as an
event generator. The goal is to check how the reaction plane reconstruction
methods react in a well defined situation. In this test, only the most central 
collisions have been used. The flow
parameters for the different particle types are not exactly zero but close to
zero, and differ from a particle type to another.}
The effect
of {the addition of the $P_x^a$ component} 
is to add the value of $F$ to the original value of the flow parameter, 
\mod{giving a $F_{init}$ value.}
\modif{This procedure will allow to study the effects of the different reaction
plane reconstruction methods on flow parameters measurements, as in
\cite{Sul90}.}


\subsection{\textbf{One plane per event or one plane per particle?}}

Since the transverse momentum is used both for the reaction plane 
estimation and for the projection, auto-correlations are expected 
{(see full circles in figures \ref{f:pxa_vs_yr_1} and 
\ref{f:pxa_vs_yr_2})}. {Auto-correlation} effects
are also amplified by the {loss} of information due to 
\modif{a non perfect} detection. 
The usual way to solve this
problem is to remove the particle which has to be projected from the estimation
of the reaction plane. Thus, an additional corrective component 
$\overrightarrow{P^{cor}_j}$ is added to the momentum {of the
remaining particles in order to \mod{ensure the} 
conservation of the total momentum.} Its definition is the following:

\begin{equation}
 \overrightarrow{P^{cor}_j}=-\frac{A_j}{\sum_{k=1,k\ne i}^{M}{A_k}}\times 
\overrightarrow{V_i}
\end{equation}

\noindent where $i$ is the removed particle, $\overrightarrow{V_i}$ its 
velocity, $A_j$ the mass of particle $j$ \mod{($j \neq i$)} 
and $M$ the event multiplicity. In
this case, one plane is determined for each particle of the event.

Two prescriptions have { been} used for {the methods described in section
\ref{desc_meth}}. In the first
one, one plane per event { is} determined. This allows to check the effect of
the auto-correlations with respect to the method used. In the second
one, one plane per particle { is} determined, {in order to test the 
efficiency of the correction.} 

\subsubsection{Azimuthal dispersion between the true reaction
plane and the reconstructed one}

To compare the relative efficiencies of these methods, the distribution
of the angular azimuthal difference $\Delta\Phi$ between
the true and the reconstructed reaction plane directions has been studied.
\mod{Such distributions have been already shown (see figure 6 of \cite{Ang97}).

The observed mean value $<\Delta\Phi>$ is zero.} 
{The accuracy of the reaction plane determination is estimated with 
the standard deviation $\sigma(\Delta\Phi)$ of these distributions as 
a function of the \modif{added} flow parameter value $F$. 
\modif{Figure \ref{f:delta_phi} shows such evolutions of 
$\sigma(\Delta\Phi)$ with $F$. $F$ has been used instead of \Finit because the
\Finit values are different from a particle type to another, whereas the same
$F$ value has been added to all particles.}
The upper row corresponds to 
the Ar + Ni system, the lower row, to the Xe + Sn system. The left column
correspond to the case when one plane per particle is determined and the right
column to the one plane per event prescription. No significant difference is
found between these two prescriptions}. 
For all methods and systems, the dispersion $\sigma(\Delta\Phi)$ 
decreases when $F$ increases. \modif{For small $F$ values, the different methods
give slightly different results because the initial flow parameter is not zero}. As expected, the
reaction plane determination is more accurate in case of strong in-plane
emission. For both systems, the transverse momentum method and the tensor method
give similar accuracies, whereas for the azimuthal correlation method the
dispersion is systematically higher for all $F$ values. For the Xe + Sn
system, the values of $\sigma_{\Delta\Phi}$ are smaller than for the Ar + Ni
system. 

From these observations, three main conclusions can be {drawn}: 
{i) the
azimuthal dispersion of the reconstructed reaction plane is the same for 
the one
plane per particle and one plane per event prescriptions}; 
ii) {similarly to the conclusion of reference \cite{Ang97}}, the
azimuthal correlation method is less accurate than the {two other} 
methods, even
at low \Finit values; iii) the higher the mass of the system and the higher the
\Finit value, the more accurate is the reaction plane reconstruction.

\mod{Surprisingly, no difference is seen on 
$\sigma(\Delta\Phi)$ between the one plane per particle and the one plane per
event prescriptions, whereas strong discrepancies are seen for 
the \pxpsa=f(\Yr)
curves (figures \ref{f:pxa_vs_yr_1} and \ref{f:pxa_vs_yr_2}). One has to
\modif{keep in mind} that in the present case, only the deviation from the true reaction plane
is studied. For the \pxpsa=f(\Yr) curves, the combined effects of the
reaction plane reconstruction accuracy and of the projection of transverse
momenta on the reconstructed reaction plane are present.}

\subsubsection{Measured flow parameter versus initial flow parameter}

For the transverse momentum method, 
the dependence of \Fexp \modif{on} the initial value of the flow parameter is shown
in figure \ref{f:dan_odyn} in the case of one plane per particle 
{(left column) and in case of one plane per event (right column)}. 
{For one plane per particle, the procedure used} to remove
auto-correlations seems to be efficient for { $H$ isotopes}. For 
{heavier particles}, \Fexp is systematically below \Finit. This
underestimation increases with increasing charge. For {a given
\Finit value}, the underestimation is lower for Xe+Sn system than for 
Ar+Ni. Finally, the amplitude of this underestimation decreases with 
increasing value of \Finit.

{With one plane per event (right column of figure 
\ref{f:dan_odyn}), \Fexp values are systematically larger than \Finit
values due to the {auto-correlation} effects mentioned above. 
For a fixed \Finit value, the overestimation increases with
decreasing charge of the particle. Here again, the amplitude of the 
overestimation {diminishes} for higher values of \Finit and with 
increasing mass of the system.} 

For the momentum tensor method, results are shown in figure \ref{f:ten3} 
for one plane per particle {(left column) and for one plane per 
event (right column)}. The same trends are observed as for the
transverse momentum method. The {auto-correlation}
effects are \mod{smaller} in the case of one plane per event. 

For the azimuthal correlation method {(figure \ref{f:wilson})}, 
the same trends are observed as for the two other methods. {The
main difference is a somewhat larger underestimation of the flow 
parameter value.}

\subsubsection{Influence of the experimental set-up}

Part of the anti or auto {correlation} may be related to the detector
efficiency. In order to probe the effect of the experimental set-up, \modif{the 
 INDRA ``filter''} has been applied on the 
generated events. \modif{The INDRA ``filter'' is a sophisticated software which
simulates the response of the detector. For each particle,
the energy losses in each layer of the detector are calculated. 
An identification procedure
similar to the experimental one is
then applied on the energy losses giving back the charge and the energy of the
detected particle. Doing this, multiple hits in a
detector are treated in the same way they are treated in the experiment. This
procedure allowed to reproduce the angular and energy thresholds
observed experimentally.}

The reaction plane is then
reconstructed from the so called ``filtered'' events. 
In figure \ref{f:flot_filt}, the correlation between \Fexp and \Finit 
is plotted for the momentum tensor method with the one plane per 
\mod{event} prescription. For the two systems studied here, the results 
are very similar to those obtained with a perfect detection (see the
right column of figure \ref{f:ten3}). The measured flow values are 
above the initial ones for the Ar + Ni system, and close to the 
initial ones for the Xe + Sn system. The detector has a weaker effect 
than the reaction plane determination procedure. This conclusion is 
identical to those made in reference \cite{Sul90} 
{for a $4\pi$ array which had
higher thresholds}. 

\subsubsection{Conclusions of the simulations}

The following conclusions can be drawn from the previous study: \\
i) removing the particle of interest from the reaction plane
introduces an anti-correlation
which is not {counterbalanced} by adding a corrective momentum 
$\overrightarrow{P^{cor}_j}$ to the momentum of other particles. The 
anti-correlation effect leads to an underestimation of sideward flow values;\\ 
{ ii) when one plane per particle is determined,
the amplitude of the anti-correlation is higher for small \Finit values, for
heavy particles and for small system size;}\\
iii) { when one plane per event is determined}, 
the amplitude of the auto-correlation is higher for small \Finit values, for
light particles and for small system size;\\
iv) { for all methods, for the one plane per event prescription,}
the auto-correlation effect {leads to an overestimation of the flow 
parameter value.}
{ This overestimation is} minimum for the momentum tensor method;\\ 
{v) the best results are obtained for the momentum tensor method 
with the one plane per event prescription;}\\ 
{vi) Detection effects are weaker than effects induced by the
procedures used to reconstruct the reaction plane.}

{One could conclude from these simulations that the best method 
to measure small sideward flow values is the momentum tensor method
with one plane per event. Unfortunately, the simulation is too simple 
compared to the experimental situation. We remind the reader that in 
this simulation, the mid-rapidity contribution is only present
for protons and neutrons. This is why the ``auto-correlation''
effect is mainly seen for protons in the simulation. In the
experimental data, the mid-rapidity emission is also made of heavier 
particles \cite{Lefort00,Dore98}. The autocorrelation 
effect is seen for all particles for the Ar + Ni system (full circles 
in figures \ref{f:pxa_vs_yr_1} and \ref{f:pxa_vs_yr_2}) as well as for 
the Xe + Sn system (full circles in figure \ref{f:flot_xesn}). In the 
latter case, the flow parameter is weakly dependent on the incident 
energy. Therefore, from the experimental results, it turns out that 
the ``auto-correlation'' effects are strong. In other words, 
qualitative effects may be understood thanks to the simulations but 
quantitative estimations of the corrections required in the data are 
difficult to realize from these simulations. Such 
a correction may be done if the ``auto-correlation'' effect is well 
understood. The study of the origins of the ``auto-correlation'' 
effect is done in the next section.}

\subsection{\textbf{Origin of the auto-correlations}}

{ These studies} show us that the methods developed at high energies, { where
high values of sideward flow are measured,} 
are not well suited for intermediate energies where the flow parameter 
values are typically around or below 30 MeV/c/A.
The amplitude of the anti or auto correlations depends also on the nature of
the particle. Let us try nevertheless to {identify 
the origin} of the auto-correlations at intermediate energies.

First of all, one has to make some remarks. If a method 
{would be} able to reconstruct
perfectly the reaction plane from the momentum of all particles, no
auto-correlation effect would be seen. This is shown for example 
{in the right column of figure \ref{f:ten3}} 
for the Xe+Sn system. The experimental flow parameter \Fexp is
very close to \Finit for \Finit values above 20 MeV/c/A, although the momenta
of all particles {have been used} to calculate the momentum
tensor. This is quite
unexpected since for high \Finit values, the transverse momenta values are
large. The auto-correlation effect should be maximum for high \Finit. 
The usual explanation of the auto-correlation effect {does} 
not seem to be the right one. If so, where does this
``auto-correlation'' effect come from ? Since the azimuthal
correlation method is less accurate than the two other ones, 
the origin of the ``auto-corellations'' will not be checked for this method.

\subsubsection{``Auto-correlations'' for the transverse momentum method}

In the transverse momentum method, one
assumes that the particles emitted above \Ycm are all coming from the decay of
the quasi-projectile (QP), and those emitted below \Ycm are all coming from the 
decay of
the quasi-target (QT). But at intermediate energies, the contributions from
the QT, the QP and from the mid-rapidity area are mixed,
especially for the most violent collisions 
\cite{Lefort00,Dore98}. { In addition, for the most central collisions, 
the azimuthal angular
distributions are rather flat and no privileged direction can be clearly seen. 
Therefore }
a wrong weight {may be} attributed to the particles, {and}
the estimated reaction plane {may have} nothing to do with the true one. 
{In this case, the reconstructed reaction} 
plane {may be oriented along} the particles with the
highest momenta. 

{On the other hand,} if the right weight was attributed to the right particles,
this effect should vanish. This can be checked in the simulation, for which the
origin of particles is known. The results are shown in figure \ref{f:flot_bin}. 
In this
simulation, a pure binary scenario has been assumed: the first stage of the
collision leads to the formation of a quasi-projectile and a quasi-target both
deflected in the reaction plane, without any pre-equilibrium {emissions}. 
One plane per event is determined, using the transverse momentum 
method, and the weight of the
particles is attributed according to their origin: +1 for the particles emitted
by the quasi-projectile, -1 for those emitted by the quasi-target. 
It is seen that the effect of ``auto-correlation'' is
\mod{removed}, even for the smaller flow values. 
The so called ``auto-correlation'' {effect comes } from a loss
of information (the origin of the detected particles), instead of the use of
the transverse momenta in both the reaction plane determination and in the
projection.

\modif{In the experiment, the exact knowledge of the origin of the particle is
impossible, especially for the most damped reactions. In addition 
the collisions are not purely binary due to prompt emissions.
The promptly emitted particles carry a part of the total tansverse
momentum and the QP and QT are hence pushed out of the reaction plane.
In the experiment, the perfect determination of the reaction plane using the
tranverse momentum method is very difficult, 
especially for the central collisions.}

\subsubsection{``Auto-correlations'' for the momentum tensor method}

For the momentum tensor method, the origin of ``auto-correlations'' effects 
can be understood using a simple test. Let us consider the case where
a quasi-projectile of mass number $A$ splits {into} two equal
size fragments and the quasi-target of the same mass number $A$
remains unchanged. {The quasi-projectile and the quasi-target are
deflected in the reaction plane. The axis joining their center of mass  
has an angle $\theta_{deflec}$ with respect to the beam axis 
(see figure \ref{f:schema_test_tenseur})}. The axis joining 
the two fragments issued from the splitting of the quasi-projectile 
has an angle $\theta_{split}$ with respect to the beam direction and 
an angle $\phi_{split}$ with respect to the reaction plane. 
The relative velocity between the quasi-projectile and the
quasi-target is $V_r$ and the relative velocity between the two 
fragments of the quasi-projectile is $\alpha V_r$. A scheme of the 
described {configuration} is presented in figure 
\ref{f:schema_test_tenseur}. 

{The momentum tensor of such a simple case can be calculated and
one can study the azimuthal angular difference $\rm{\Delta}\phi$ 
between the reconstructed reaction plane and the true one.
The direction of the reaction plane is defined in the plane
transverse to the beam axis, displayed in the right 
panel of figure \ref{f:schema_test_tenseur}. In this panel, the QT is 
by definition in the reaction plane but the two splitting 
fragments are out of the plane. 

\mod{The first key variable is $\alpha$. 
When its value is small (close to zero), the QP-QT axis
is obviously the main axis and the reaction plane is perfectly determined. But
when the $\alpha$ value is big enough 
the axis between the two
splitting fragments may become the main axis. In this case, the
reconstructed reaction
plane direction may be dependent on the 
splitting direction $\phi_{split}$ which is the second key variable.} 
 
In figure \ref{f:test_tenseur}, the evolutions of $\rm{\Delta}\phi$ 
as a function of $\alpha$ and $\phi_{split}$ are displayed. The angles 
$\theta_{deflec}$ and $\theta_{split}$ are set respectively to 
10\degre~ and 40\degre~. Similar pictures are obtained for other 
$\theta_{deflec}$ and $\theta_{split}$ values. 
For large enough values of $\alpha$ 
$\rm{\Delta\phi}$ is 
strongly dependent on $\phi_{split}$. For low values of $\alpha$, i.e. small 
relative velocities of the two fragments compared to the relative
velocities of the quasi-projectile and the quasi-target,
$\rm{\Delta\phi}$ does not depend on $\phi_{split}$ 
and is equal to zero. The two \mod{out-of-plane} fragments do not introduce 
much pertubations on the reconstruction procedure. The direction of 
the reconstructed reaction plane is mainly determined by the QT. 
The reaction plane is therefore well estimated. On the other hand, 
when $\alpha$ is larger than 1, the reaction plane is mainly
determined by the two out-plane fragments. The direction of the 
reconstructed reaction plane is therefore \mod{correlated} to the 
splitting direction. In the experiment, this last configuration is 
similar to the most violent (central) collisions but with a larger 
multiplicity.} 

{If the two fragments issued from the splitting of the quasi-projectile
were gathered before applying the momentum tensor method, the reaction plane
would be perfectly determined whatever the fragmenting \mod{configuration}. 
That means that if the origin of the fragments is
known, on can determine perfectly the reaction plane by grouping the fragments
coming from the same source in a single fragment. 
The results of the momentum tensor method depend on the way the fragments are 
gathered.}
\modif{As for the tranverse momentum method, a perfect determination of the
reaction plane by using the momentum tensor method is 
very difficult in the experiment.}


\subsection{\textbf{Discussion}}

{For incident energies around the balance energy, the standard
methods used for the reaction plane reconstruction are not well suited.
Whatever the method used, the one plane per event prescription leads
to an ``auto-correlation'' effect i.e. a large overestimation of the 
flow parameter values. At variance, the one plane per particle 
prescription induces an anti-correlation effect which gives an
underestimation of the flow parameter.}

This indicates that the negative flow values observed in experimental data can 
be attributed to \mod{the used reconstruction methods}, 
especially for the heaviest fragments. For the
light charged particles, the physical effect can not be completely 
\mod{ruled out},
since positive values of flow are observed for alphas, whereas negative values
are expected if the reaction plane determination methods effects are dominant.
For such particles, the two effects are probably mixed and more detailed
studies have to be performed to establish their relative weights in the
observed values. {To obtain the true flow parameter values, one has to
understand the auto-correlation effects in order to correct them accurately.}

{But understanding the origin of the ``auto-correlation'' effects is a
complicated task. To correct
them, a complete 
knowledge of the origin of particles is needed. This can be
achieved only for the less violent collisions and/or at higher energies where
the mixing between the different contributions is weak. For the most violent
collisions, such a knowledge is unreachable unless assumptions are made. But in
this case, the flow values obtained may only result from these assumptions.}

Since the correction of experimental data seems to be impossible, it could be
easier to apply \mod{the experimental filter and the
{analysis} procedure on theoretical calculations}. 
Most of the available dynamical
calculations have to evolve
to enable this procedure, since most of them are following the time evolution
of the one body density. \modif{More precisely, the dynamical calculations 
should include the proper description of particle and fragment formation.}

\section{Conclusions}

The in-plane flow parameter has been determined for the Ar + Ni collisions from
32 A.Mev to 95 A.Mev, for the Ni + Ni system from 32 A.Mev to 90 A.MeV 
{and for the Xe + Sn system from 25 A.MeV to 50 A.MeV. For
central collisions, the balance energies are equal to 82 
$\rm{\pm}$ 2 A.MeV for the Ar + Ni system and 75 $\rm{\pm}$ 2 A.MeV
for the Ni + Ni system. For the Xe + Sn system, the balance energy is around 
50 A.MeV, but experiments at higher incident energies have to be
performed to confirm the result}. These values are in agreement with
the systematics of balance energies found in other experiments. \mod{This
systematics has been reproduced by a dynamical model assuming 
\Kinf$\approx$~220 MeV and
\snn=0.8~\snnf \cite{DLM92}}. As already
observed, the balance energy
weakly depends on the \mod{particle} nature. For these central collisions, 
negative
flow values are observed for both systems. {Two different explanations are
proposed for these negative values. 

{The first one, supported by transport model calculations, attributes 
this effect to the relative importance between the prompt emission 
and the evaporative one. A negative flow parameter value can be observed 
if the prompt emission is dominant. This explanation seems to be satisfactory
for the light particles. Deuterons, tritons and $^3He$ are 
predominantly emitted in the mid-rapidity regions
\cite{Lefort00,Dore98} and their flow parameter values are \mod{indeed} 
negative.
At variance, for protons and alpha particles, the measured flow parameters are 
positive. They are emitted as much by the QP and QT \mod{as at 
mid-rapidity}.
\mod{But on the other hand}, negative flow values are measured for fragments
while the prompt emission is not the dominant process for these 
products. \mod{For these heavy fragments, 
the observed negative values cannot be
explained in this way.}}  

The second explanation attributes these negative values to the 
experimental methods used to extract the reaction plane. The usual 
method used to avoid auto-correlations, {the omission of the particle
of interest,} leads to an \mod{anti-correlation}. This induces 
an inversion of the reaction plane reaction and then lead to 
the measurement of negative flow values. {The amplitude of
this effect increases when the flow parameter value decreases i.e. 
when the incident energy is getting closer to the balance energy.}
This explanation is supported by the observation of negative flow
values for the heaviest fragments, whereas a positive value is
expected. A careful study of the ``auto-correlation''
effect shows that \modif{its manifestation results from the loss of 
information about the product
origins for both methods.} 

In experimental data, these two effects are probably \mod{mixed up}. 
They disturb the
measurement of the absolute value of sideward flow, 
{especially around the balance energy for which low flow
parameter values are expected}. On the other hand, the
relative evolutions with incident energy, and especially the 
determination of the balance energy, are in agreement with previous 
experimental studies and theoretical calculations. 
It may indicate a relative robustness of the balance energy variable.
In the present status,} 
the real effect can only be studied with simulations on which
the complete experimental procedure can be applied. {An accurate determination
of the in-medium nucleon-nucleon interaction parameters can only be achieved 
if the
disturbances induced by the {analysis} methods and the experimental set-up are
explicitely taken into account. This requires an evolution of dynamical calculation
to make \mod{possible this comparison procedure}.}


\newpage
\figart{ht}{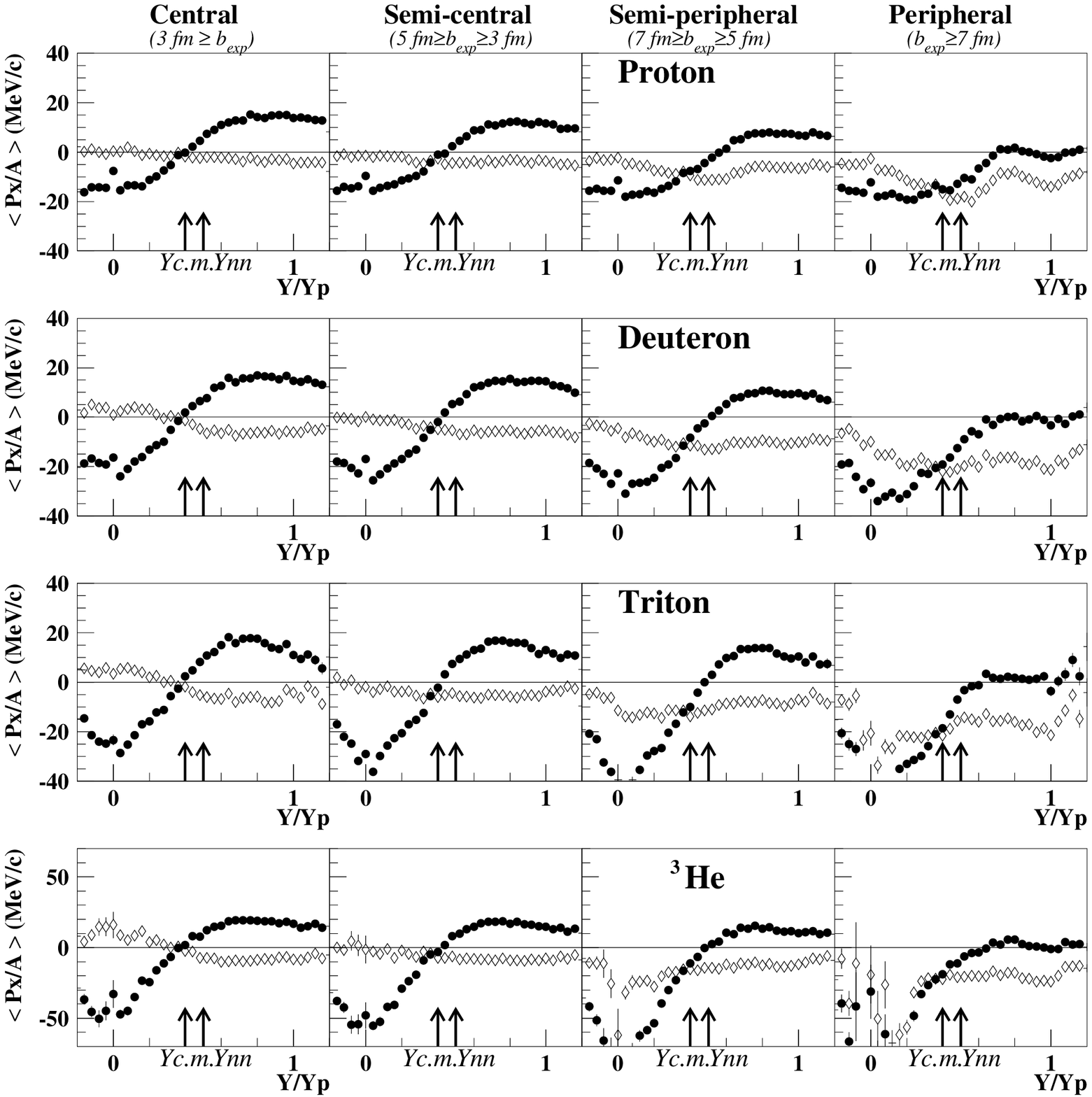}
{Variations of \pxp as a function of the reduced rapidity \Yr for the Ar + Ni
collisions at 74 A.MeV. The first row corresponds to protons the
second to deuterons the third to tritons and the fourth to $^3He$. 
Each column
corresponds to an experimental impact parameter bin, ranging from central (left
column) to peripheral collisions (right column). The open circles
correspond to the one plane per particle prescription, the full circles, to
the one plane per event prescription.}
{f:pxa_vs_yr_1}{0.7}

\figart{ht}{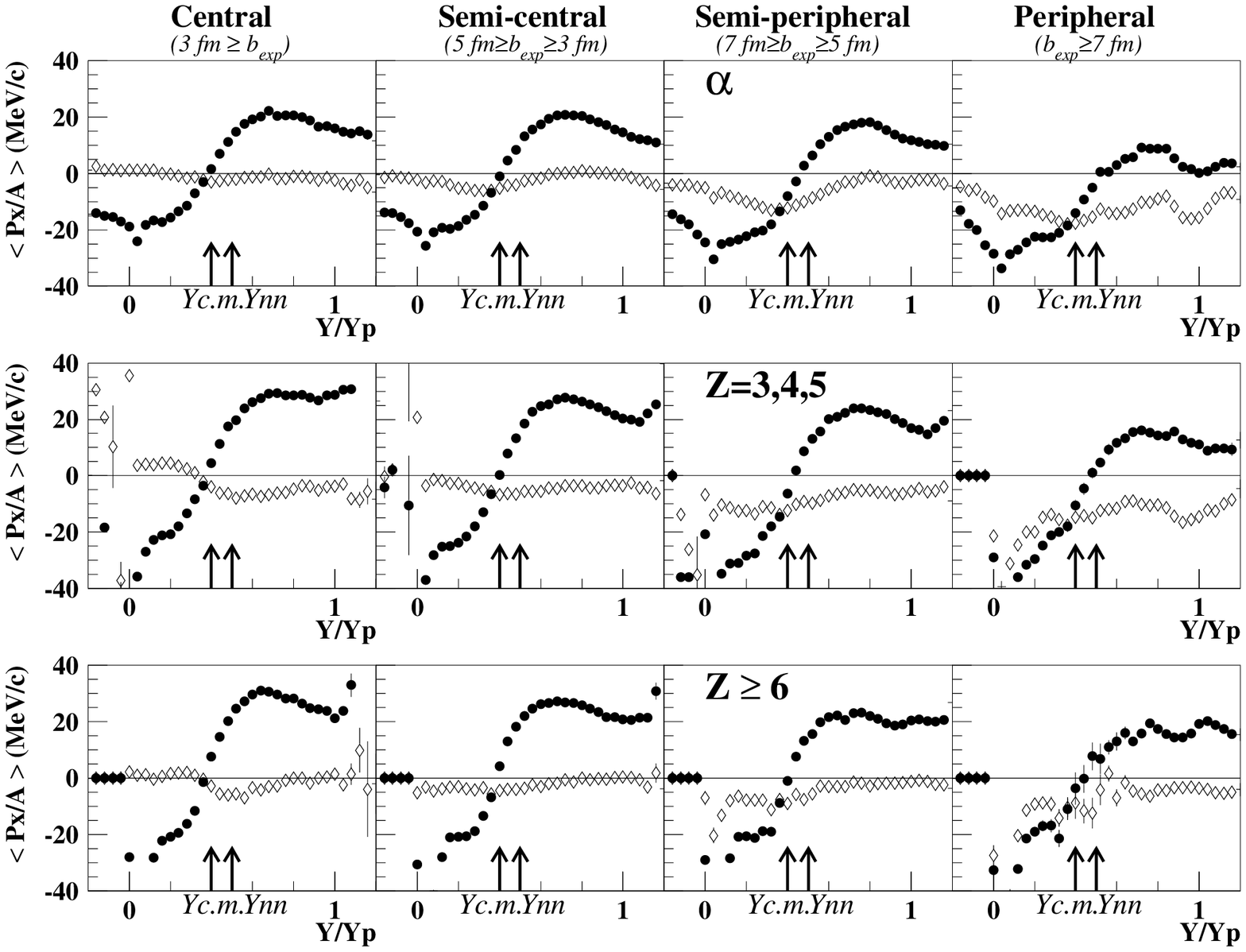}
{Same as \ref{f:pxa_vs_yr_1} for
alphas (first row) for Z=3,4,6 (second row) and for Z$\ge$6 (third row).}
{f:pxa_vs_yr_2}{0.7}

\figart{ht}{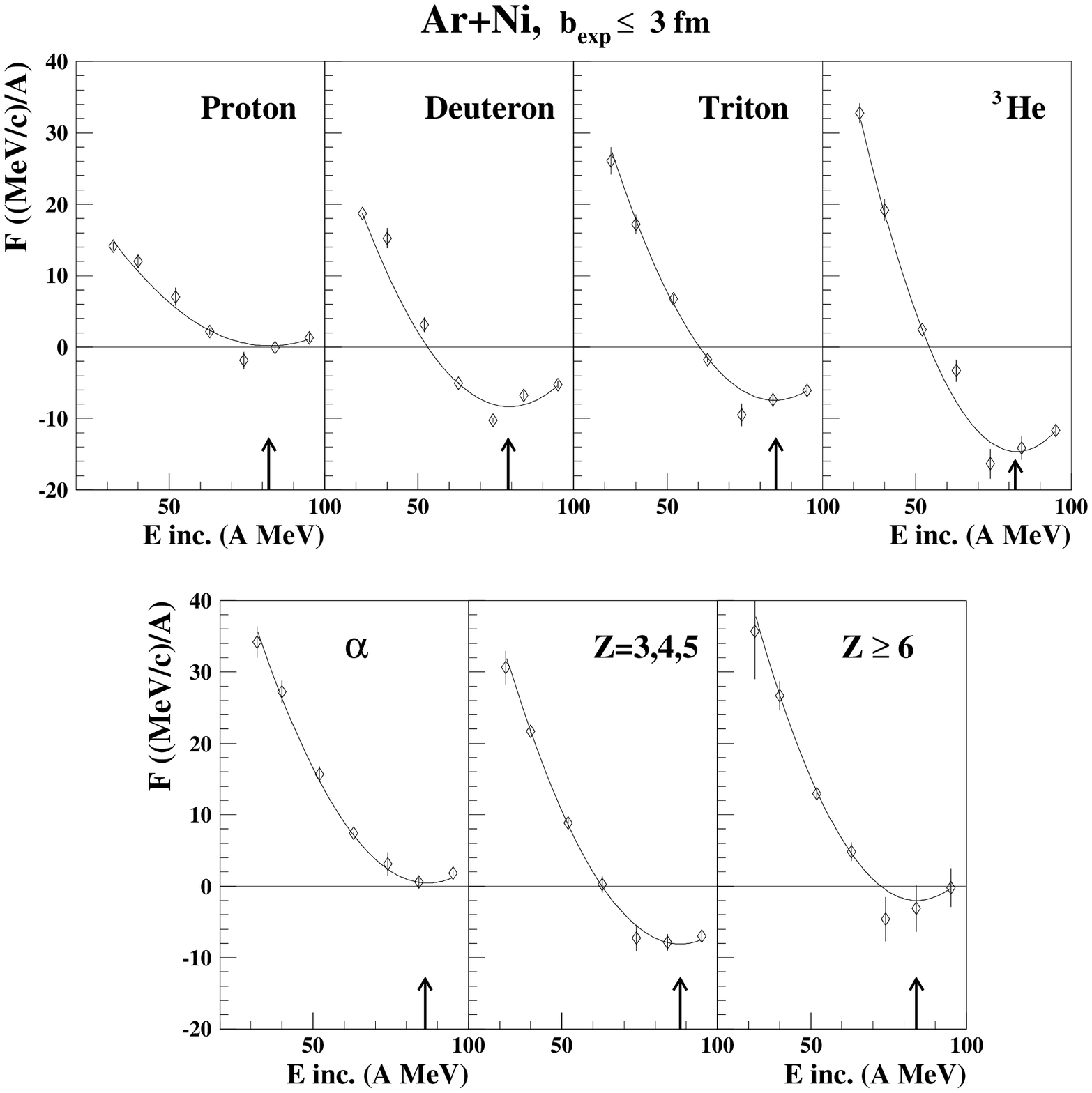}
{Variations of the flow parameter $F$ with the incident
energy for the most central Ar + Ni collisions (\bexp $\le$ 3 $fm$) and for
different particle natures. On each spectrum, the line is a quadratic
fit \mod{to} the
experimental value. The balance energy \Ebal corresponding to the
minimum $F$ value is indicated by an arrow. \mod{The momentum tensor method with the
one plane per particle prescription has been used to reconstruct the reaction
plane.}}{f:flot_arni}{0.6}

\figart{ht}{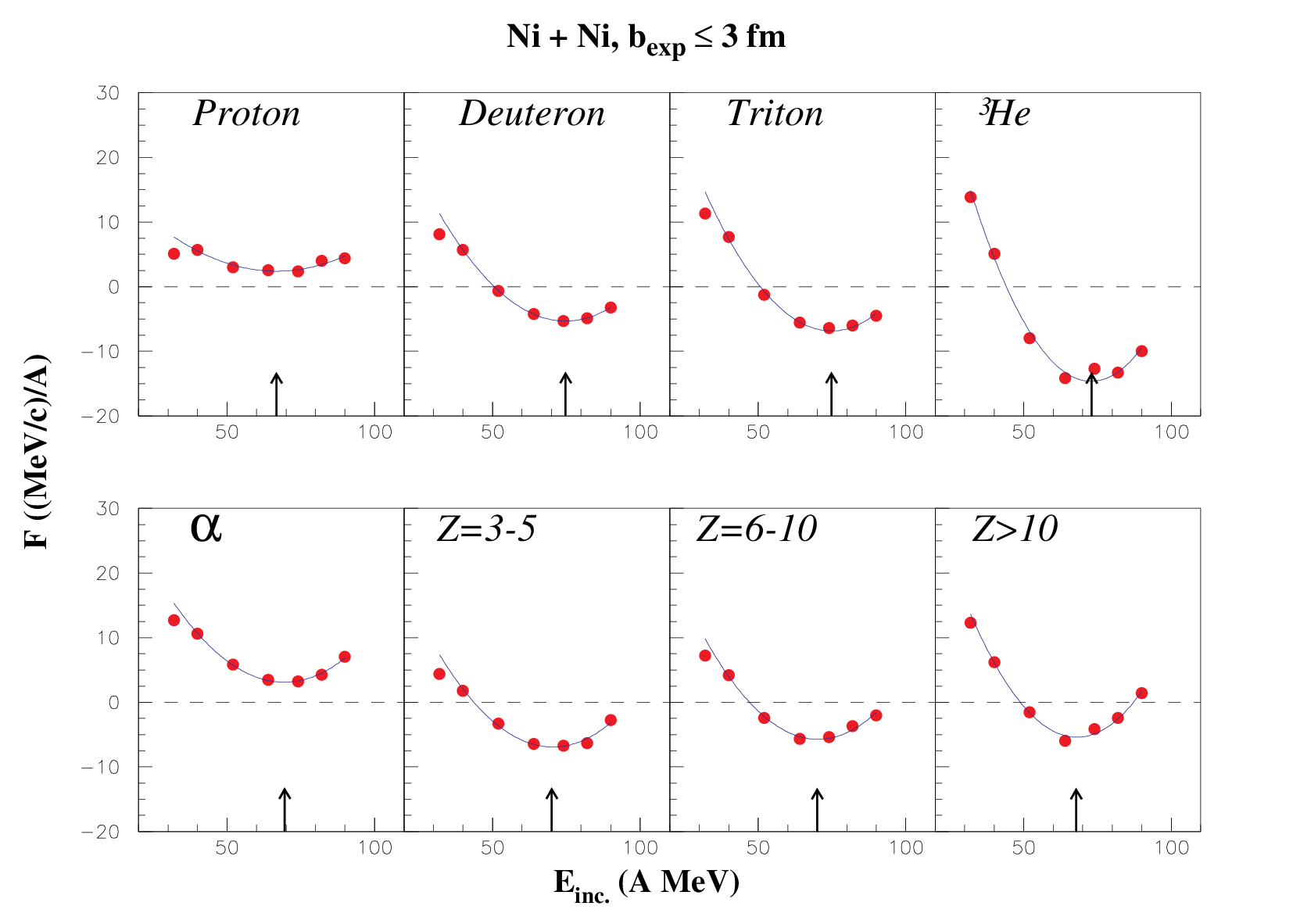}
{Variations of the flow parameter $F$ with the incident
energy for the most central Ni + Ni collisions (\bexp $\le$ 3 $fm$) and for
different particle natures. On each spectrum, the line is a quadratic fit
\mod{to} the
experimental value. The balance energy \Ebal, corresponding to the minimum $F$
value, is indicated by an arrow. \mod{The momentum tensor method with the
one plane per particle presciption has been used to reconstruct the reaction
plane.}}{f:flot_nini}{0.6}

\newpage
\figart{ht}{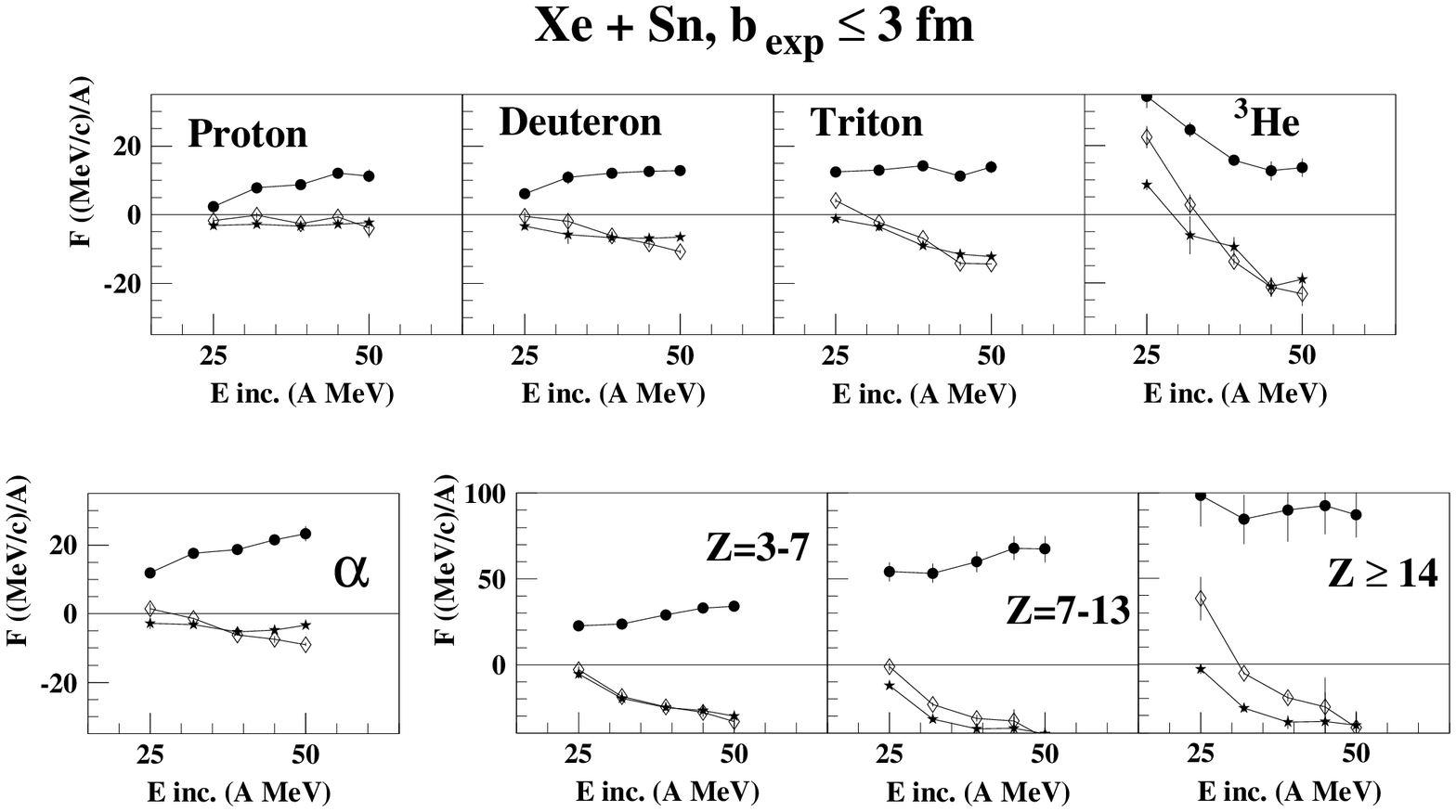}
{Variations of the flow parameter $F$ with the incident
energy for the most central Xe + Sn collisions (\bexp $\le$ 3 $fm$) and for
different particle natures. On each panel, the results of two
reaction plane determination are shown: the momentum tensor method (open
diamonds) and the transverse momentum method (stars). For both cases, one plane
per particle is determined. The full circles
correspond to the momentum tensor method when one plane per event 
is determined.}{f:flot_xesn}{0.7}

\figart{ht}{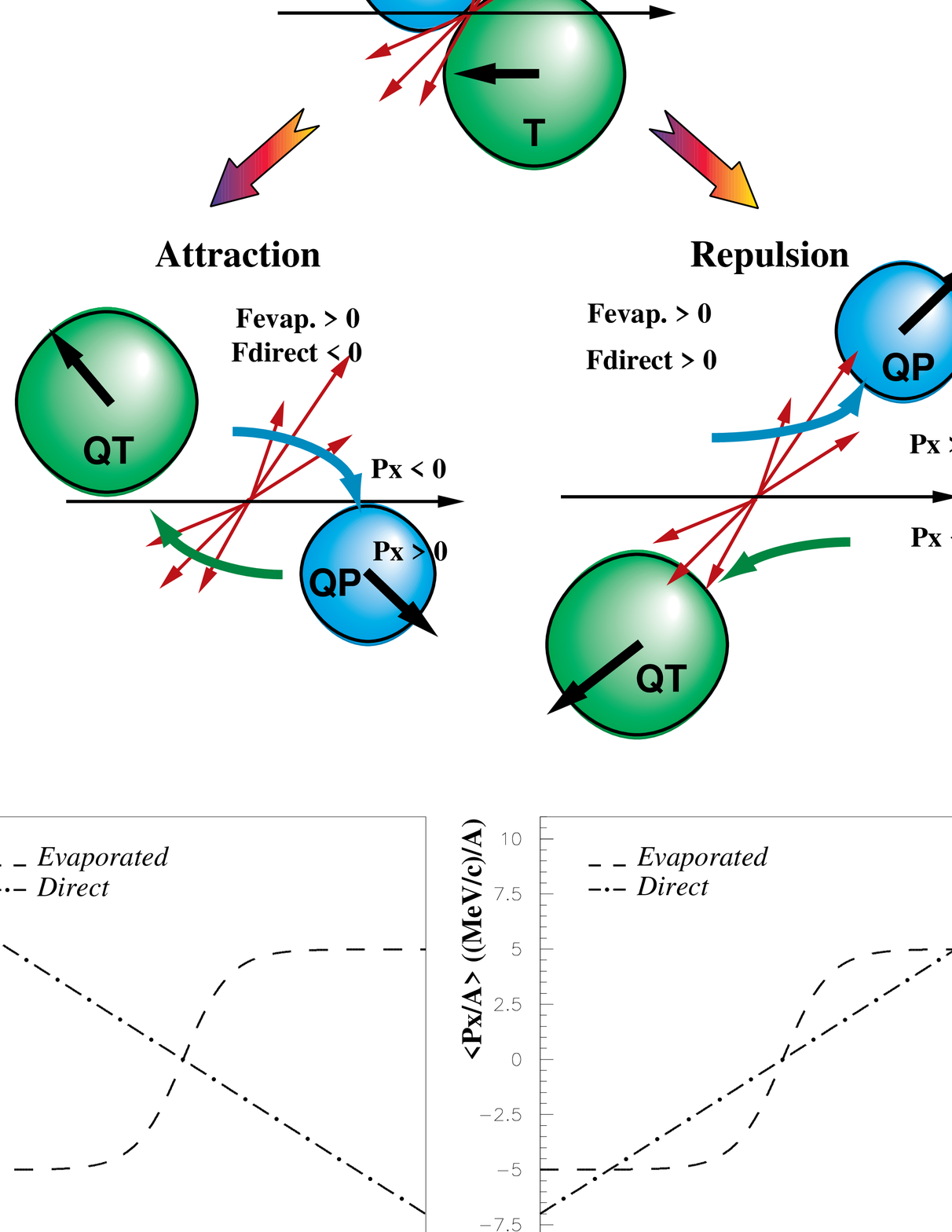}
{Schematic explanation of the two
possible contributions to the flow parameter. The uppermost picture 
correspond to {the beginning of the reaction when the projectile P} and 
the target T are touching each other. Arrows correspond to the
velocity of ``direct'' (promptly emitted) particles. {Middle pictures 
correspond to the time at which the QP and QT are leaving each other.}
The lowermost pictures correspond to the expected
variation of \pxpsa with \Yr for ``evaporated'' particles (dashed line) and 
``direct'' particles (dashed and dotted line) in case of attraction (left
column) and bounce-off (right column).}{f:flot_contribs}{0.5}

\figart{ht}{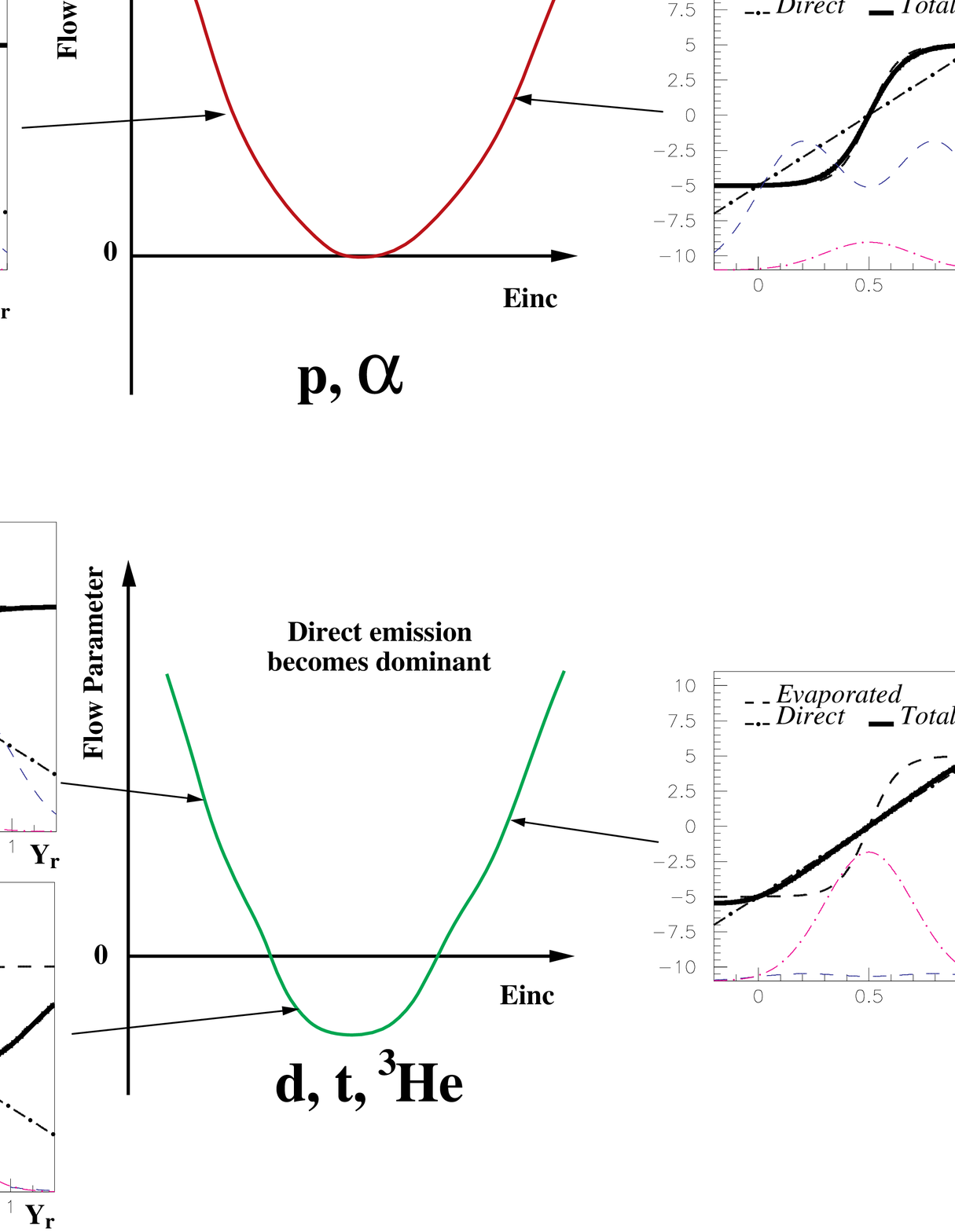}
{Expected variations of flow parameter $F$ with 
incident {energy}. {In the upper panel, the contribution of 
evaporated 
particles is always {dominant}. In the lower panel, the contribution of
``direct'' particles becomes dominant} with increasing energy.}
{f:flot_neg}{0.45}

\figartdeuxcol{ht}{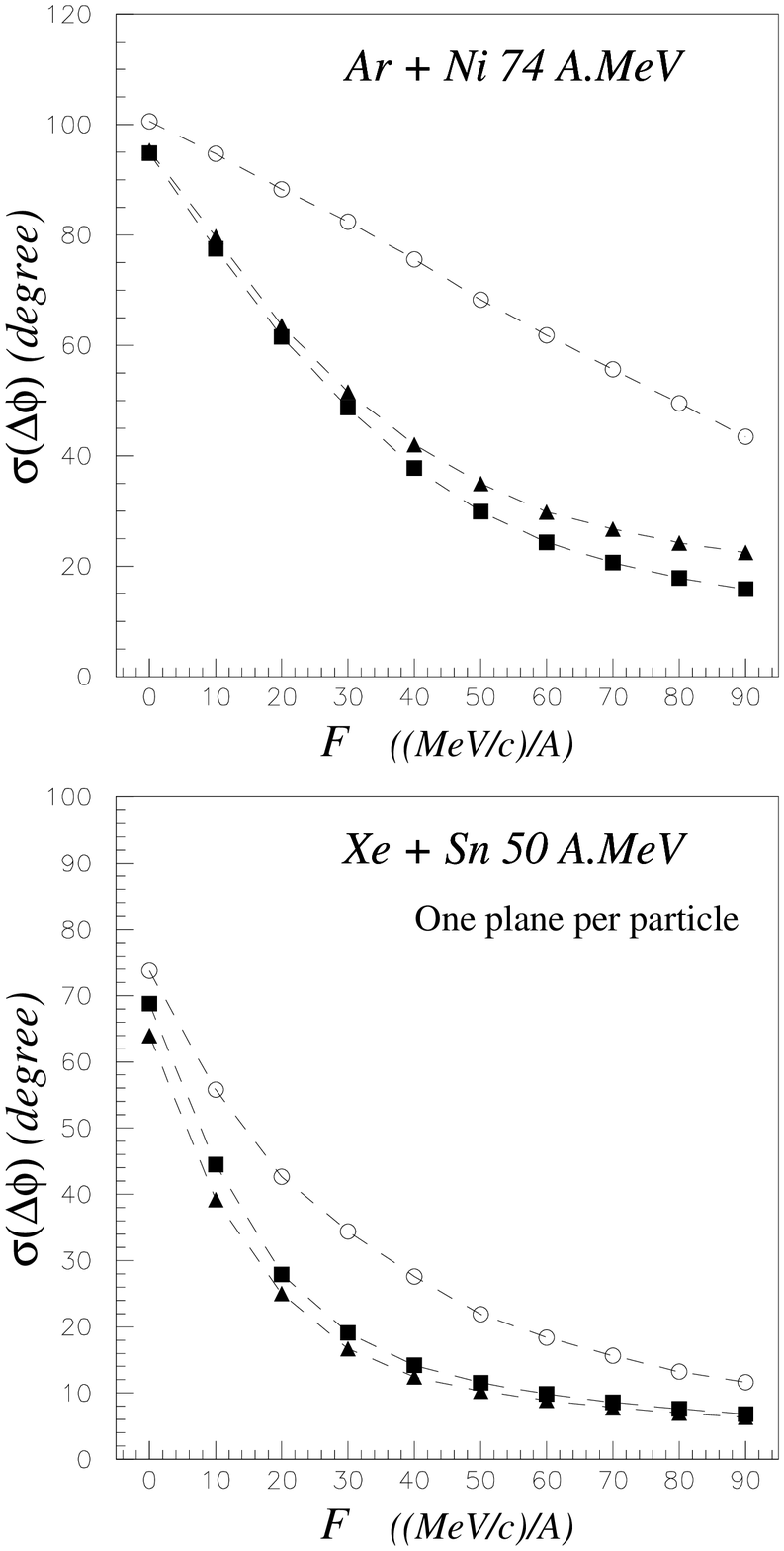}{0.5}{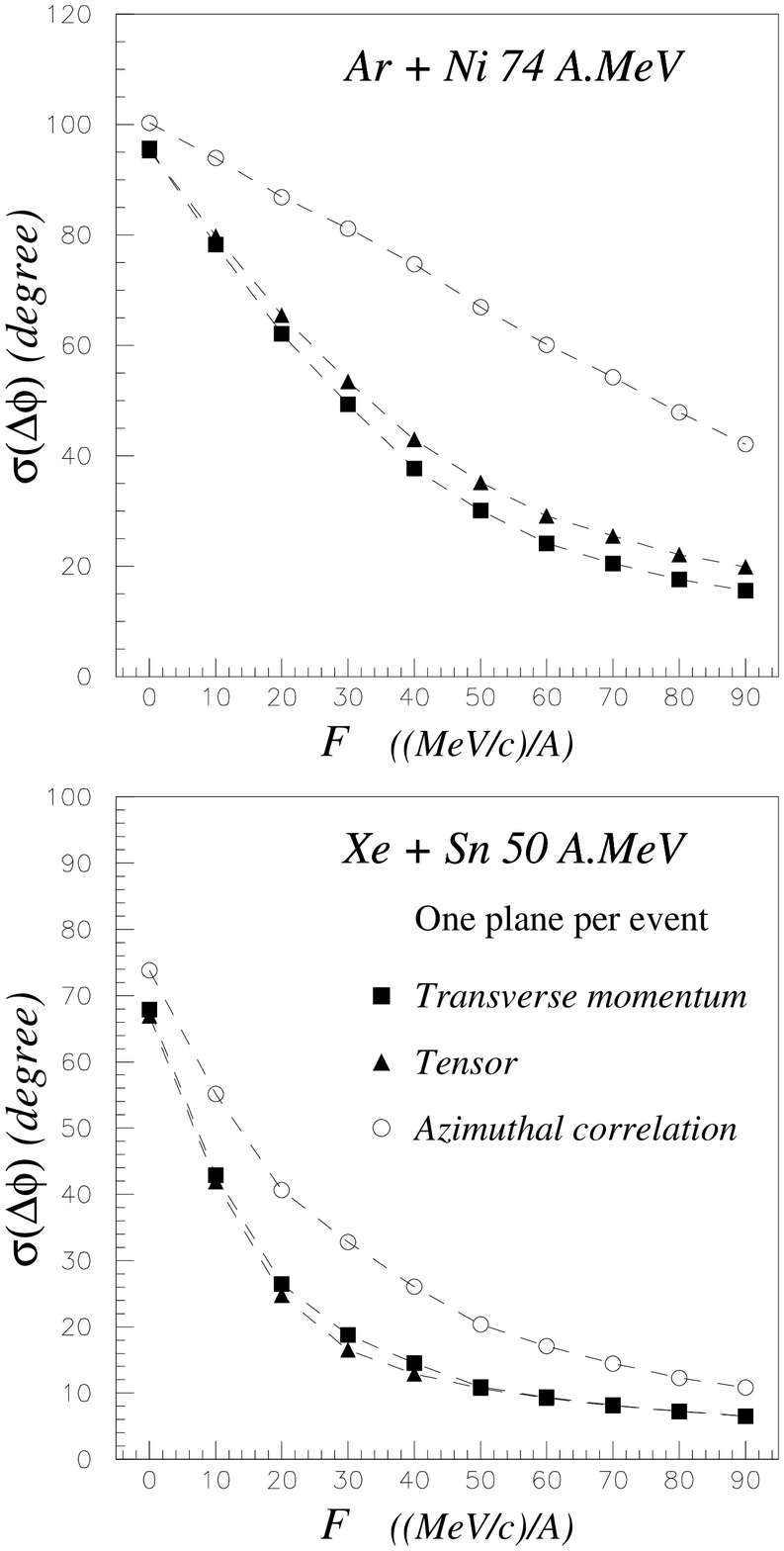}{0.5}
{Accuracy on the
reaction plane estimation obtained as a function of the initial flow parameter
value \Finit. This was obtained by using the SIMON code for the Ar + Ni system 
at 74 A.Mev {(upper row)} and for the Xe + Sn system {at 50 A.MeV}
{(lower row)}. {The left column corresponds to the one plane per
particle prescription and the right column to the one plane per event
prescription.} The 
{squares}
correspond to the transverse momentum method, the triangles to the momentum
tensor method and the {open circles} 
to the azimuthal correlation method.}
{f:delta_phi}

\figartdeuxcol{ht}{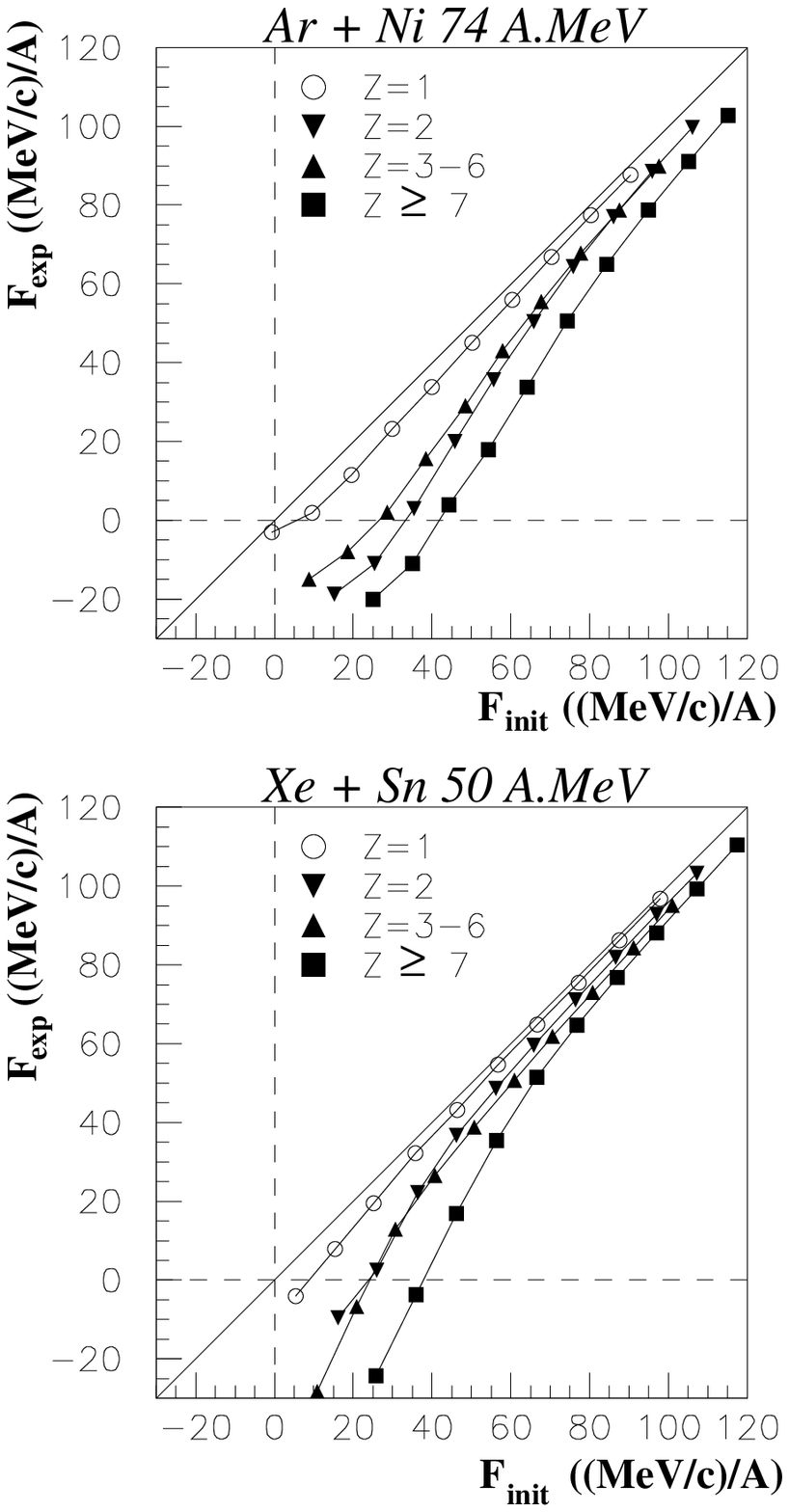}{0.7}{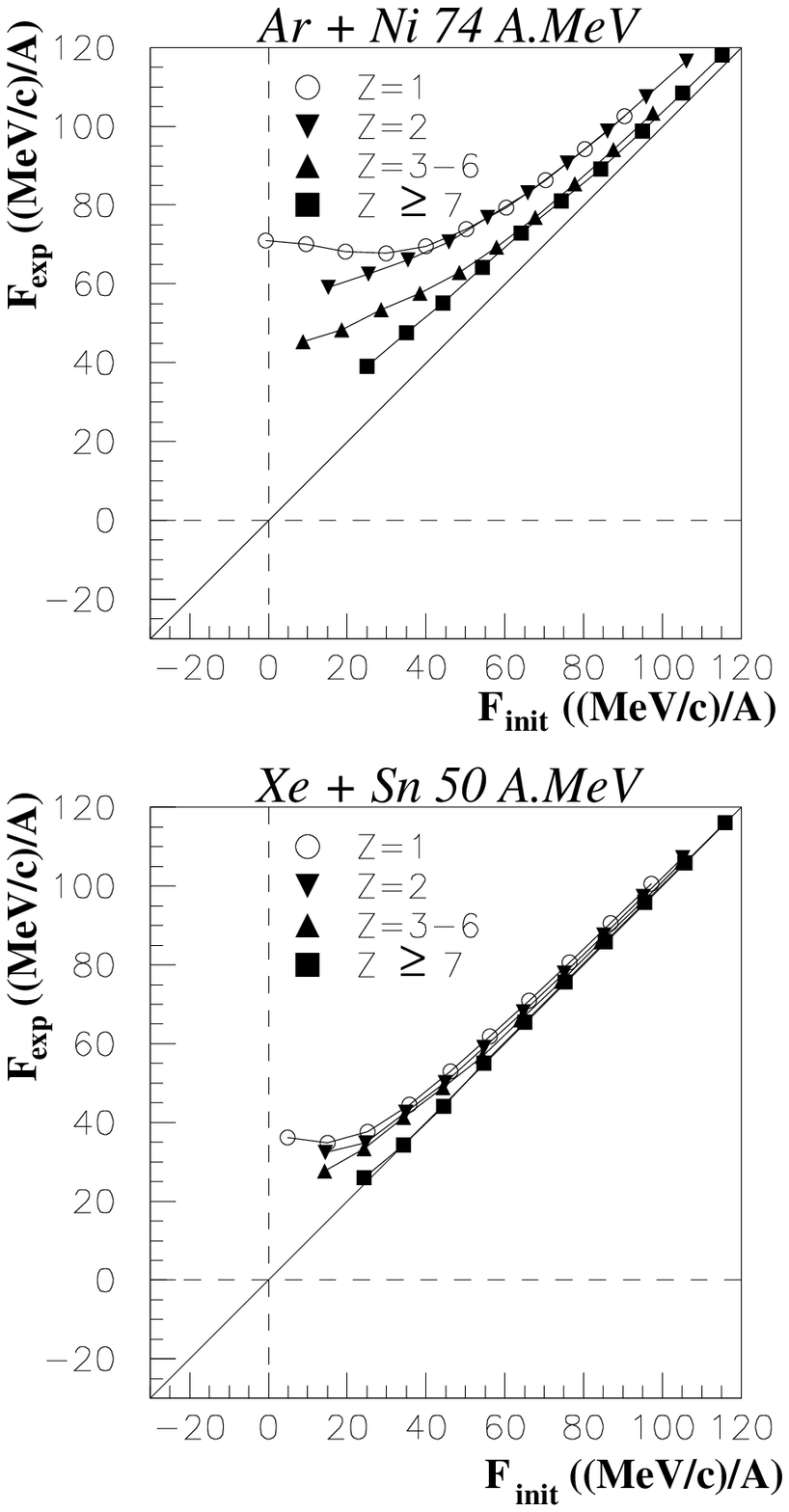}{0.7}
{Correlation of the measured flow parameter value 
\Fexp and the initial flow parameter value \Finit in the framework of the SIMON
code. The transverse momentum method has been used and one plane per particle
has been calculated. The upper panel {corresponds} to the Ar + Ni system at 74
A.MeV and the lower panel to the Xe + Sn system at 50 A.MeV. 
\mod{The left column corresponds to the one
plane per particle presciption and the right column to the one
plane per event presciption.}}
{f:dan_odyn}

\figartdeuxcol{ht}{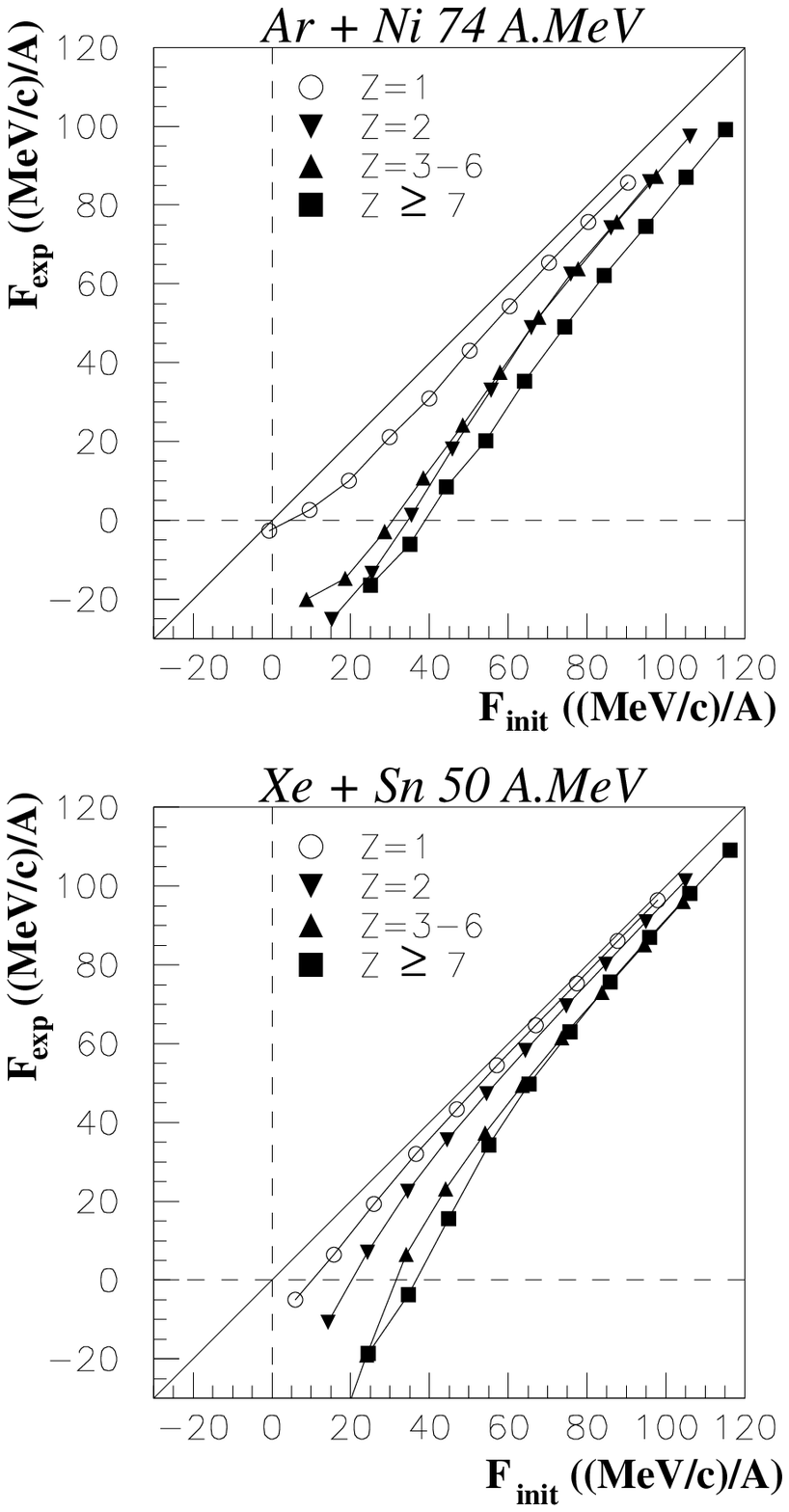}{0.7}{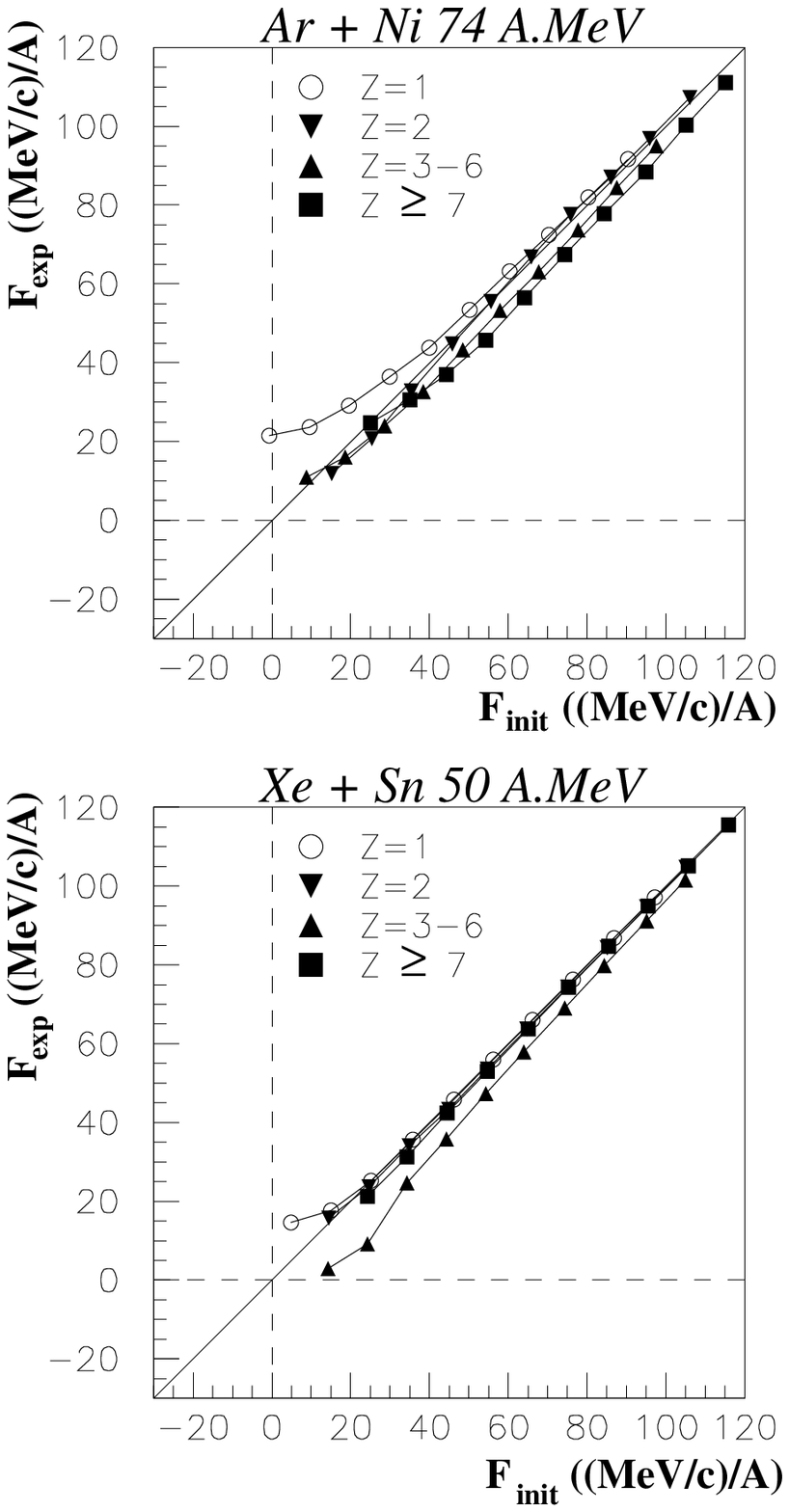}{0.7}
{Same as \ref{f:dan_odyn} by using the momentum
tensor method.}{f:ten3}

\figartdeuxcol{ht}{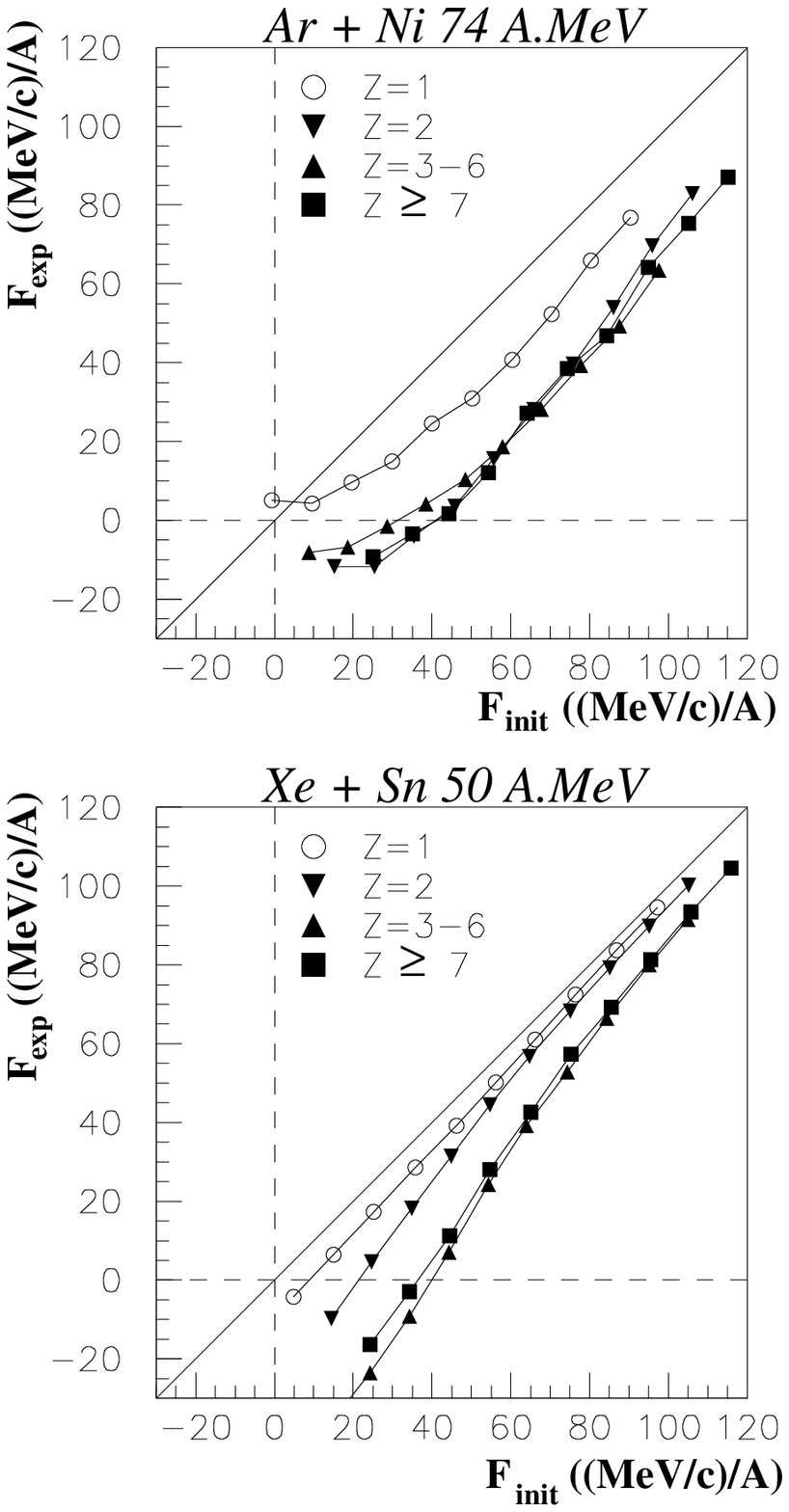}{0.7}{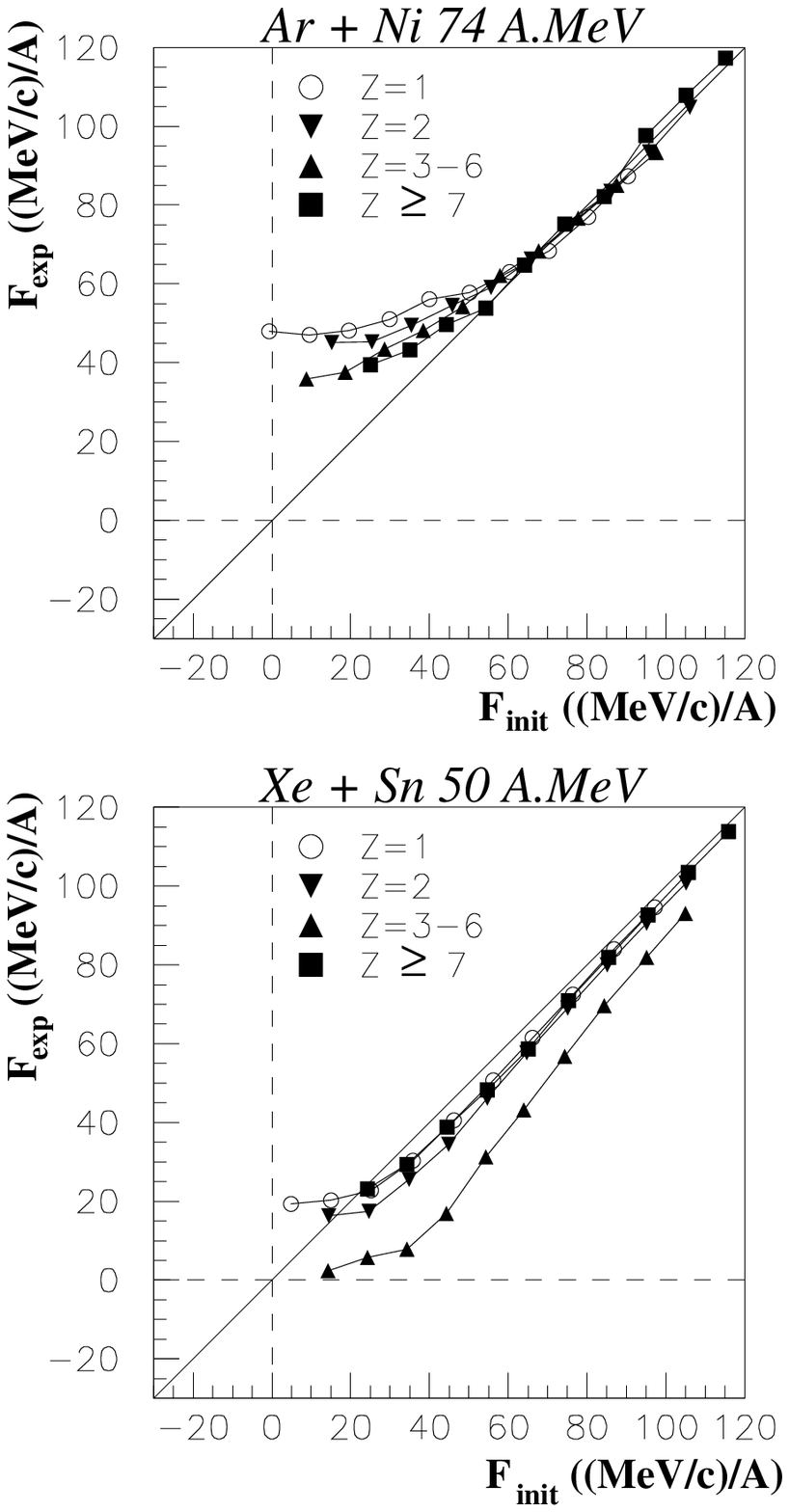}{0.7}
{Same as \ref{f:dan_odyn} by using the azimuthal
correlation method.}
{f:wilson}


\figart{ht}{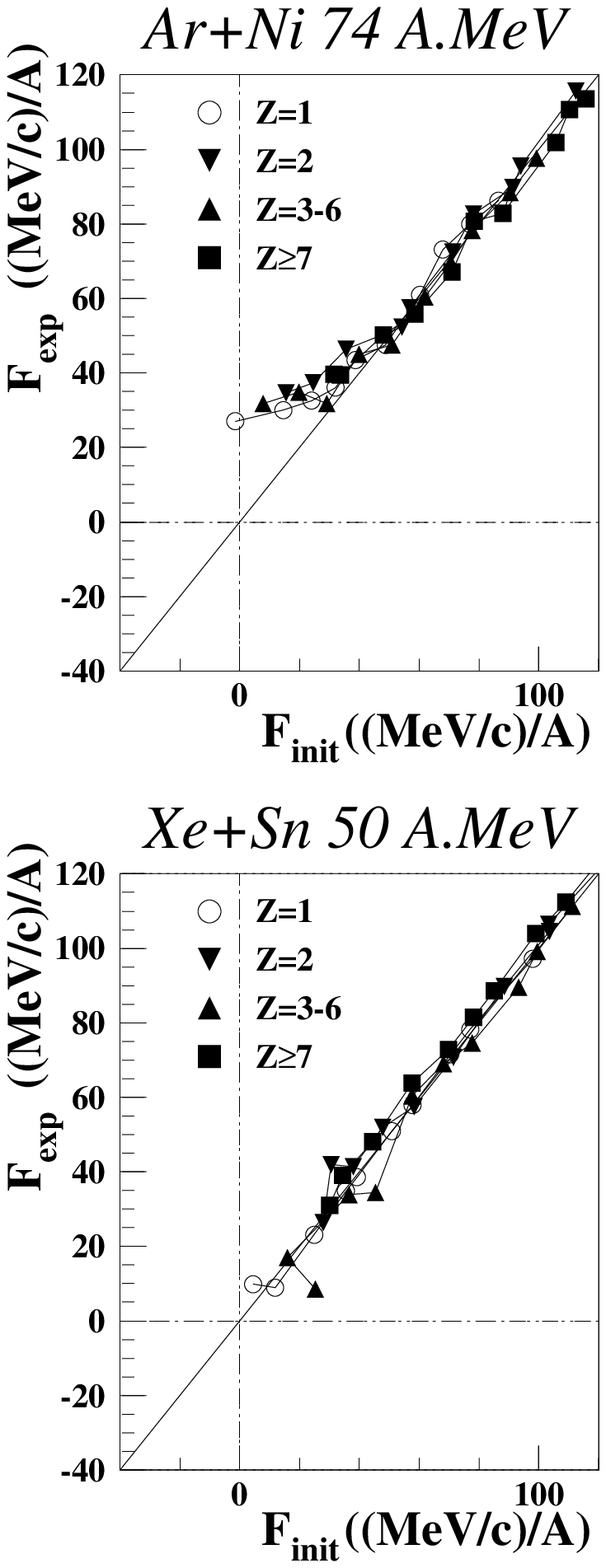}
{Same as {\ref{f:ten3} right column}
but the events have been affected
by the INDRA {filter}.}{f:flot_filt}{0.7}

\figartdeuxcol{ht}{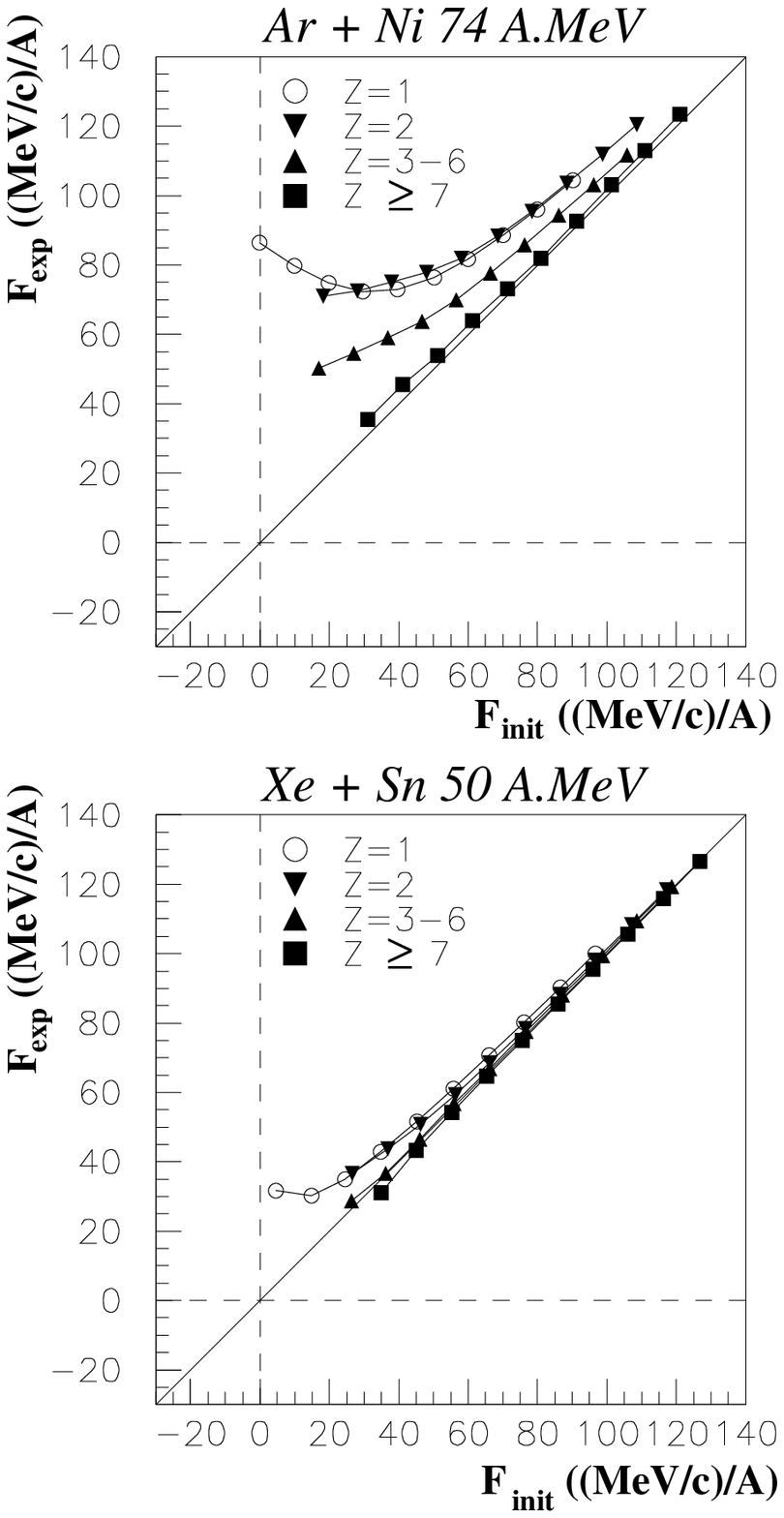}{0.7}{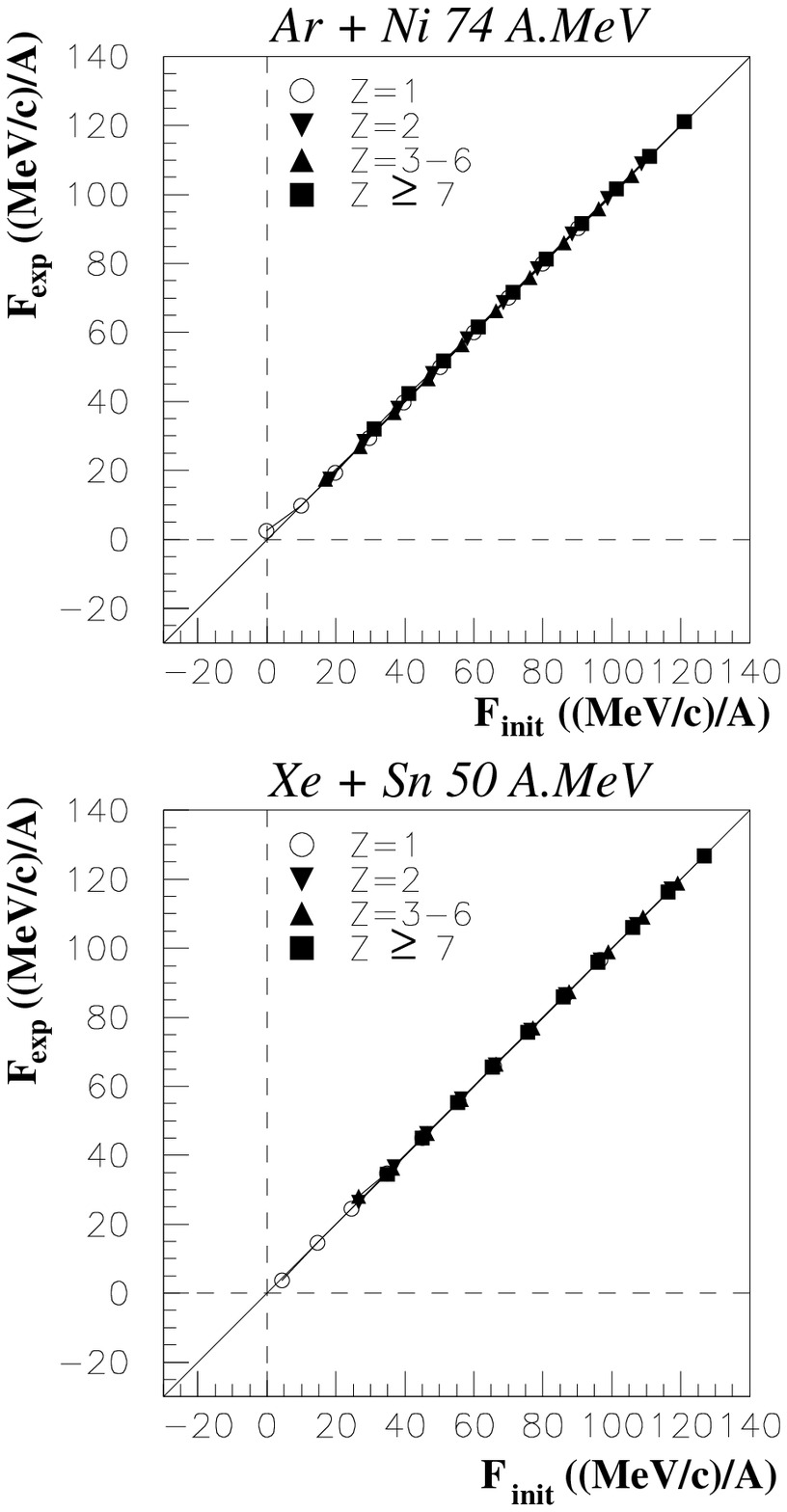}{0.7}
{Correlation of the measured flow 
parameter value 
\Fexp and the initial flow parameter value \Finit in the framework of the SIMON
code assuming a pure binary scenario (excited QP and QT only).
{Left column: the transverse momentum method} has been used and the weights are
determined according to the reduced rapidity of the particle. 
{Right column: the weights are set
according to the true origin of the particle.} 
One plane per event has been calculated}{f:flot_bin}

\figartdeuxcolvar{ht}
{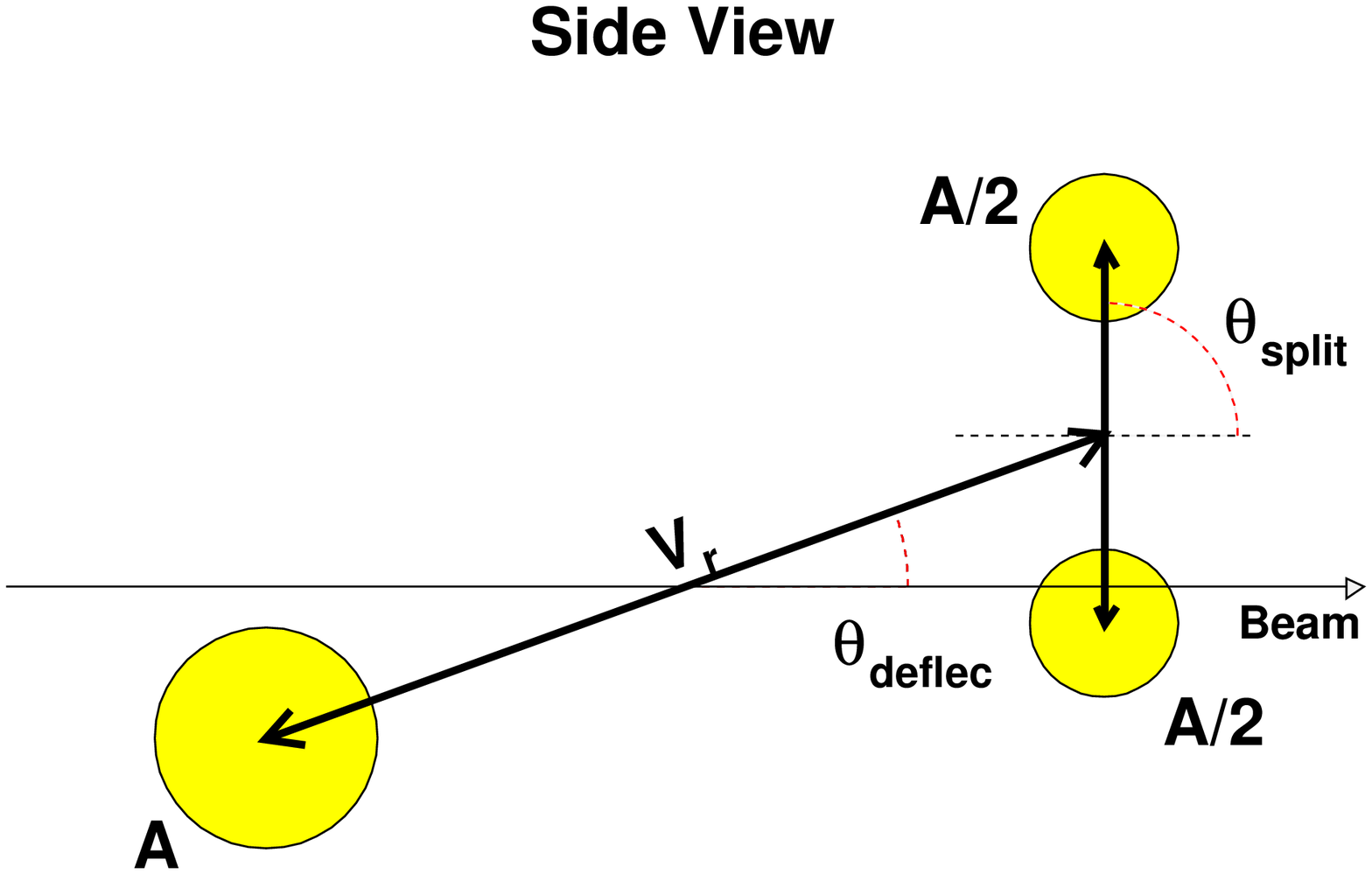}{9.8 cm}{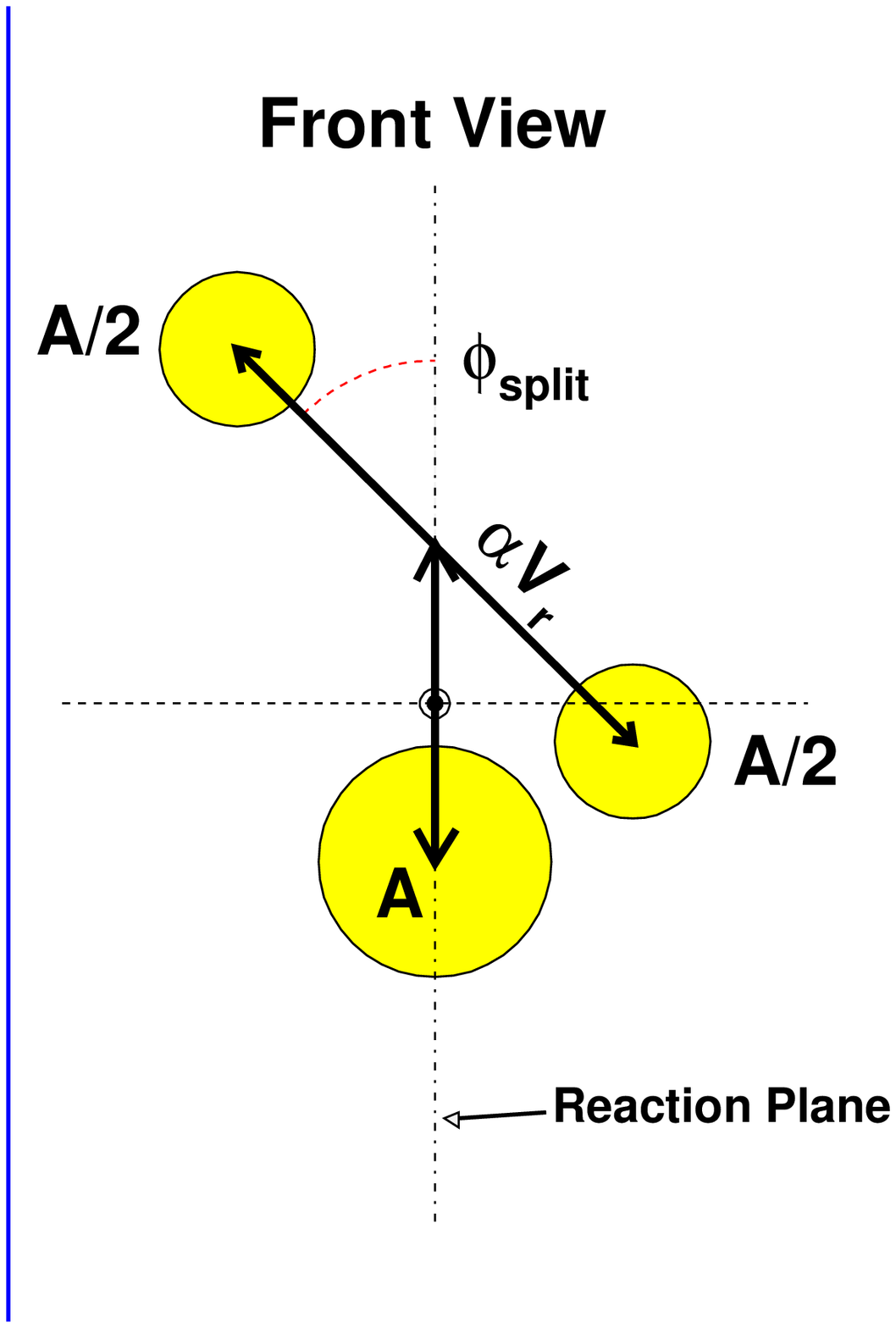}{9.8 cm}
{Scheme of the configuration used to test the ``auto-correlation''
effect for the momentum tensor method (see
text).}{f:schema_test_tenseur}{0.65}{0.25}

\figart{ht}{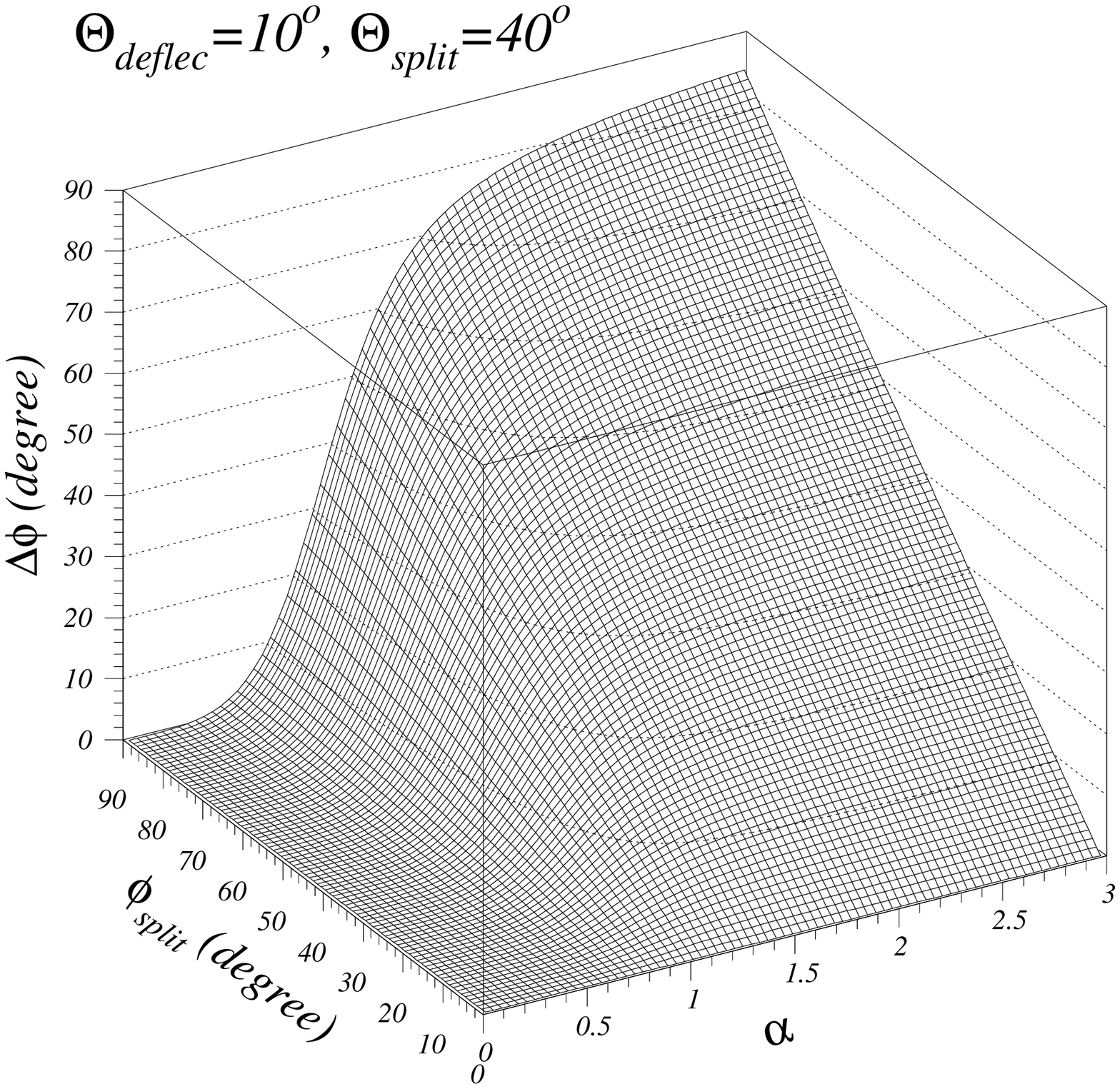}
{Variations of the angle $\phi$ between the true
reaction plane and the estimated one by using the momentum tensor method
{for $\theta_{deflec}=10^{\circ}$ and $\theta_{slpit}=40^{\circ}$}. See
text for details.}{f:test_tenseur}{0.8}

\end{document}